\documentclass[twocolumn,aps,prb,showpacs,longbibliography,10pt]{revtex4-2} 

\usepackage[normalem]{ulem}
\usepackage{amsfonts,amssymb,amsmath,graphicx,color}
\usepackage{comment}
\usepackage{array,multirow}
\usepackage{mathtools}
\usepackage{enumitem}
\usepackage{braket}
\usepackage{physics}
\usepackage[version=4]{mhchem}
\usepackage{float}
\usepackage[colorlinks=true,linkcolor=blue,citecolor=blue,urlcolor=blue]{hyperref}

\newcommand{\ie}{{\it i.e.~}} 	
\newcommand{\etc}{{\it etc.}} 	
\newcommand\commentout[1]{}

\newcolumntype{P}[1]{>{\centering\arraybackslash}p{#1}}

\definecolor{greenPR}{rgb}{0.00, 0.6, 0.00} 

\begin{document}
	
		\title{Berry Curvature and Quantum Metric in $N$-band Systems -- an Eigenprojector Approach}
	\author{Ansgar Graf}
	\author{Fr\'ed\'eric Pi\'echon}
	\affiliation{%
		Universit\'e Paris-Saclay, CNRS, Laboratoire de Physique des Solides, 91405, Orsay, France\\
	}%
	\date{\today}
	
	\begin{abstract}
The eigenvalues of a parameter-dependent Hamiltonian matrix form a band structure in parameter space. In such $N$-band systems, the quantum geometric tensor (QGT), consisting of the Berry curvature and quantum metric tensors, is usually computed from numerically obtained energy 
eigenstates. 
Here, an alternative approach to the QGT based on eigenprojectors and (generalized) Bloch vectors is exposed. It offers more analytical insight than the eigenstate approach. In particular, the full QGT of each band can be obtained without computing eigenstates, using only the Hamiltonian matrix and the respective band energy. Most saliently, the well-known two-band formula for the Berry curvature in terms of the Hamiltonian vector is generalized to arbitrary $N$. The formalism is illustrated using three- and four-band multifold fermion models that have very different geometrical and topological properties despite an identical band structure. From a broader perspective, the methodology used in this work can be applied to compute any physical quantity or to study the quantum dynamics of any observable without the explicit construction of energy eigenstates.
	\end{abstract}
	
	\maketitle

\section{Introduction}
\label{sec:intro}

In a quantum mechanical system characterized by an $N$-dimensional Hilbert space, the $N\times N$ (Hermitian) Hamiltonian matrix $H(\mathbf{x})$ frequently depends on a set of parameters $\mathbf{x}=(x_1,x_2,...)$. In condensed matter physics, such parameters may take the form of a crystal momentum, the intensity of an external field, an applied strain or any mean-field order parameter. A Hamiltonian of this kind has (real) eigenvalues $E_\alpha(\mathbf{x})$, and eigenstates $\ket{\psi_\alpha(\mathbf{x})}$, where $\alpha\in\{1,...,N\}$. The eigenvalues form a band structure, and one may thus speak of an \emph{$N$-band system}, or equivalently a \emph{(parametric) SU($N$) system}.

In this paper, we will be concerned with the quantum geometric properties of $N$-band systems. 
More specifically, we will focus on a fundamental quantum geometric object,
 the (Abelian) \emph{quantum geometric tensor (QGT)} \cite{Wilczek_1989,Resta_2011,Kolodrubetz_2017}
\begin{equation}
	T_{\alpha,ij}(\mathbf{x})\equiv g_{\alpha,ij}(\mathbf{x})-\frac{i}{2}\Omega_{\alpha,ij}(\mathbf{x}).
	\label{eq1}
\end{equation}
Here, the \emph{quantum metric (tensor)} \cite{Provost_1980} $g_{\alpha,ij}(\mathbf{x})=\Re T_{\alpha,ij}(\mathbf{x})$ is symmetric in the indices $i,j$, and the \emph{Berry curvature (tensor)} \cite{Berry_1984,Simon_1983} $\Omega_{\alpha,ij}(\mathbf{x})=-2\Im T_{\alpha,ij}(\mathbf{x})$ is antisymmetric.

The concept of a Berry curvature has a long history. It popped up early on, for example quite explicitly in Blount's work \cite{Blount_1962}, and implicitly in studies on geometric phases  \cite{Ehrenberg_1949,Aharonov_1959,Pancharatnam_1956,Stone_1976}. 
 Nowadays, the Berry curvature is recognized as a key quantity for explaining many fundamental physical phenomena \cite{Thouless_1982,Resta_2011,Nagaosa_2010,Xiao_2010,Qi_2011,Sinova_2015}. 
The influence of the quantum metric on physical effects is more subtle but has started to attract growing attention in recent years. On the theory side, many measurable effects influenced by it were identified \cite{Zanardi_2007,Srivastava_2015,Raoux_2015,Claassen_2015,Gao_2015,Piechon_2016,Julku_2016,Freimuth_2017,Lapa_2019,Gao_2019}, several measurement protocols were developed \cite{Neupert_2013,Kolodrubetz_2013,Lim_2015,Bleu_2018,Ozawa_2018b}, and recently the full QGT of a two-level system was measured directly \cite{Yu_2019,Gianfrate_2020}. 
It is now well established that the \emph{entire QGT}, not only the Berry curvature part, is essential for understanding the quantum geometric contributions to observables in $N$-band systems. 
\begin{figure}
	\centering
	\includegraphics[width=\columnwidth]{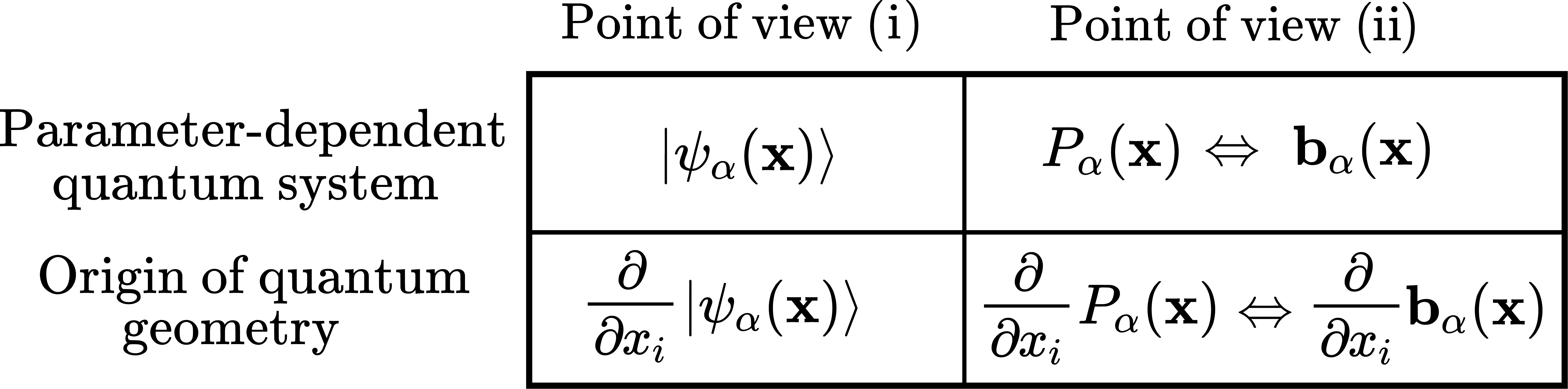}
	\caption{Two alternative points of view on quantum geometry.}
	\label{fig:pointview}
\end{figure}

The standard perspective on the QGT, following the historical development, is the following (see Fig. \ref{fig:pointview}): 
\begin{enumerate}[label=(\roman*)]
\item The system is described by the eigenstates $\ket{\psi_\alpha(\mathbf{x})}$, and the concept of quantum geometry arises from the nonzero overlap $
\bra{\psi_\alpha(\mathbf{x})}\ket{\psi_\alpha(\mathbf{x}+d\mathbf{x})}$ of eigenstates. In this language, the QGT reads \cite{Wilczek_1989,Resta_2011,Kolodrubetz_2017}
\begin{equation}
		T_{\alpha,ij}(\mathbf{x})=\bra{\partial_i\psi_\alpha}(1_N-\ket{\psi_\alpha}\bra{\psi_\alpha})\ket{\partial_j\psi_\alpha},
	\label{QG0}
\end{equation} 
where $\partial_i\equiv\partial/\partial x_i$.
\end{enumerate}
Here, our goal is to emphasize the utility of an alternative point of view that can be adopted:
\begin{enumerate}[label=(\roman*)]
\setcounter{enumi}{1}
\item The system is described by the eigenprojectors $P_\alpha(\mathbf
{x})=\ket{\psi_\alpha(\mathbf{x})}\bra{\psi_\alpha(\mathbf{x})}$. Thus, quantum geometry can be viewed as being encoded in rates of changes of the eigenprojectors upon variation of the parameters $x_i$. In this language, the QGT reads \cite{Wilczek_1989}
\begin{equation}
	T_{\alpha,ij}(\mathbf{x})=\Tr\left[(\partial_iP_\alpha)\left(1-P_\alpha\right)(\partial_jP_\alpha)\right].
	\label{QGT23}
\end{equation}
Note that there exists a one-to-one correspondence between each eigenprojector and the associated \emph{(generalized) Bloch vector} $\mathbf{b}_\alpha(\mathbf{x})$ -- the SU($N$) analog of the familiar SU($2$) Bloch vector on the unit sphere (see Section \ref{SUN}). Thus, the QGT can also conveniently be written in terms of Bloch vectors:
\begin{equation}
	\begin{aligned}
T_{\alpha,ij}(\mathbf{x})&=\frac{1}{4}\left[\partial_i\mathbf{b}_\alpha\cdot\partial_j\mathbf{b}_\alpha\right.\\
&\left.\hspace{.7cm}+i\mathbf{b}_\alpha\cdot(\partial_i\mathbf{b}_\alpha\times\partial_j\mathbf{b}_\alpha)\right].
\label{b23}
\end{aligned}
\end{equation}
\end{enumerate}
 While point of view (i) is of considerable conceptual value, eigenstates prove rather cumbersome in actual calculations of the QGT for three main reasons. First, the gauge arbitrariness in the parameter-dependent global phase of the eigenstate is problematic. Second, singularities in some of the components of $\ket{\psi_\alpha(\mathbf{x})}$ may occur at certain points $\mathbf{x}_0$ in the parameter space $X\subseteq\mathbb{R}^d$. Third, closed-form expressions for $\ket{\psi_\alpha(\mathbf{x})}$ that can be used independently of the Hamiltonian of interest are quite complicated already for $N=2$ (cf. Appendix \ref{App2states}) and essentially useless for $N>2$ \footnote{Already for three-band systems, the task of finding the QGT from energy eigenstates (in a parameter space of practical interest) can pose serious difficulties, see for instance Ref. \cite{Lim_2020}. For higher $N$ -- aside from special cases that allow for simple analytical treatment -- it is usual to resort to (not always well-controlled) numerical methods or to employ approximate perturbative analytical approaches that decouple the SU($N$) Hamiltonian into effective SU($2$) sub-Hamiltonians valid locally in parameter space.}.

Point of view (ii) is adopted more rarely. However, already the simple $N=2$ case demonstrates its usefulness for obtaining analytical insight into the relation between the system's Hamiltonian and its quantum geometry. Namely, for two-band systems, it is well known that
\begin{equation}
	\begin{aligned}
P_\alpha(\mathbf{x})&=\frac{1}{2}\left[1_2+\frac{1}{E_\alpha(\mathbf{x})}H(\mathbf{x})\right],\\
\mathbf{b}_\alpha(\mathbf{x})&=\frac{1}{E_\alpha(\mathbf{x})}\mathbf{h}(\mathbf{x}),
\end{aligned}
\label{P2b2}
\end{equation}
with $\mathbf{h}(\mathbf{x})$ a vector that decomposes the Hamiltonian into Pauli matrices [cf. Eq. (\ref{2def})]. Inserting Eq. (\ref{P2b2}) into Eq. (\ref{QGT23}) or (\ref{b23}) immediately allows to make the following observation [see Eq. (\ref{Q2}) below]: The QGT of a two-band system can be written in terms of the Hamiltonian vector $\mathbf{h}(\mathbf{x})$, the energy eigenvalues as well as their parametric derivatives. This approach circumvents the need to explicitly compute energy eigenstates. 

 The main contribution of this paper is to clarify and quantify the generalization of this convenient property to arbitrary $N$. The QGT $T_{\alpha,ij}(\mathbf{x})$ will be written in terms of only the Hamiltonian [matrix $H(\mathbf{x})$ or vector $\mathbf{h}(\mathbf{x})$] and the band energy $E_\alpha(\mathbf{x})$, without having to construct energy eigenstates at all.

Over the last few years, some efforts in this spirit have been made, mostly focusing on the $N=3$ case \cite{Barnett_2012,Lee_2015,Bauer_2016}.
Very recently, Pozo and de Juan \cite{Pozo_2020} pointed out that, quite generally, any observable (not only the QGT) can be computed without energy eigenstates if the eigenenergies are known. While similar in spirit, our work is less ambitious in scope, since we essentially restrict ourselves to the QGT. The issue of computing more complicated physical quantities without using eigenstates is deferred to future work \cite{Graf}. 
 
The setup of this paper is as follows. In Section \ref{SU2}, the familiar two-band case is reviewed. This sets the stage for the $N$-band generalization, which is developed in Sections \ref{SUN} and \ref{blochgeo}. First, in Section \ref{SUN}, we focus on the eigenprojectors and Bloch vectors, \ie we derive formulas $P_\alpha(\mathbf{x})=P_\alpha(E_\alpha(\mathbf{x}),H(\mathbf{x}))$ and $\mathbf{b}_\alpha(\mathbf{x})=\mathbf{b}_\alpha(E_\alpha(\mathbf{x}),\mathbf{h}(\mathbf{x}))$ that generalize Eq. (\ref{P2b2}) to any $N$.
Second, in Section \ref{blochgeo}, these results are combined with Eqs. (\ref{QGT23}) \& (\ref{b23}), yielding the QGT of $N$-band systems as a function of the Hamiltonian and its eigenvalues. Writing the Berry curvature in terms of the Hamiltonian vector, we recover the familiar SU($2$) result as well as the SU($3$) formula found in Ref. \cite{Barnett_2012}, and write down new explicit formulas for the SU($4$) and SU($5$) case.
Section \ref{examples} serves to illustrate the formalism using explicit multifold fermion models. In particular, we introduce a three- (four-) band low-energy model that has exactly the same spectrum as a simple pseudospin $S=1$ ($S=3/2$), but completely different geometrical and topological properties. 
Finally, we sum up and conclude in Section \ref{conclude}.

\begin{figure*}
	\centering
	\includegraphics[width=\textwidth]{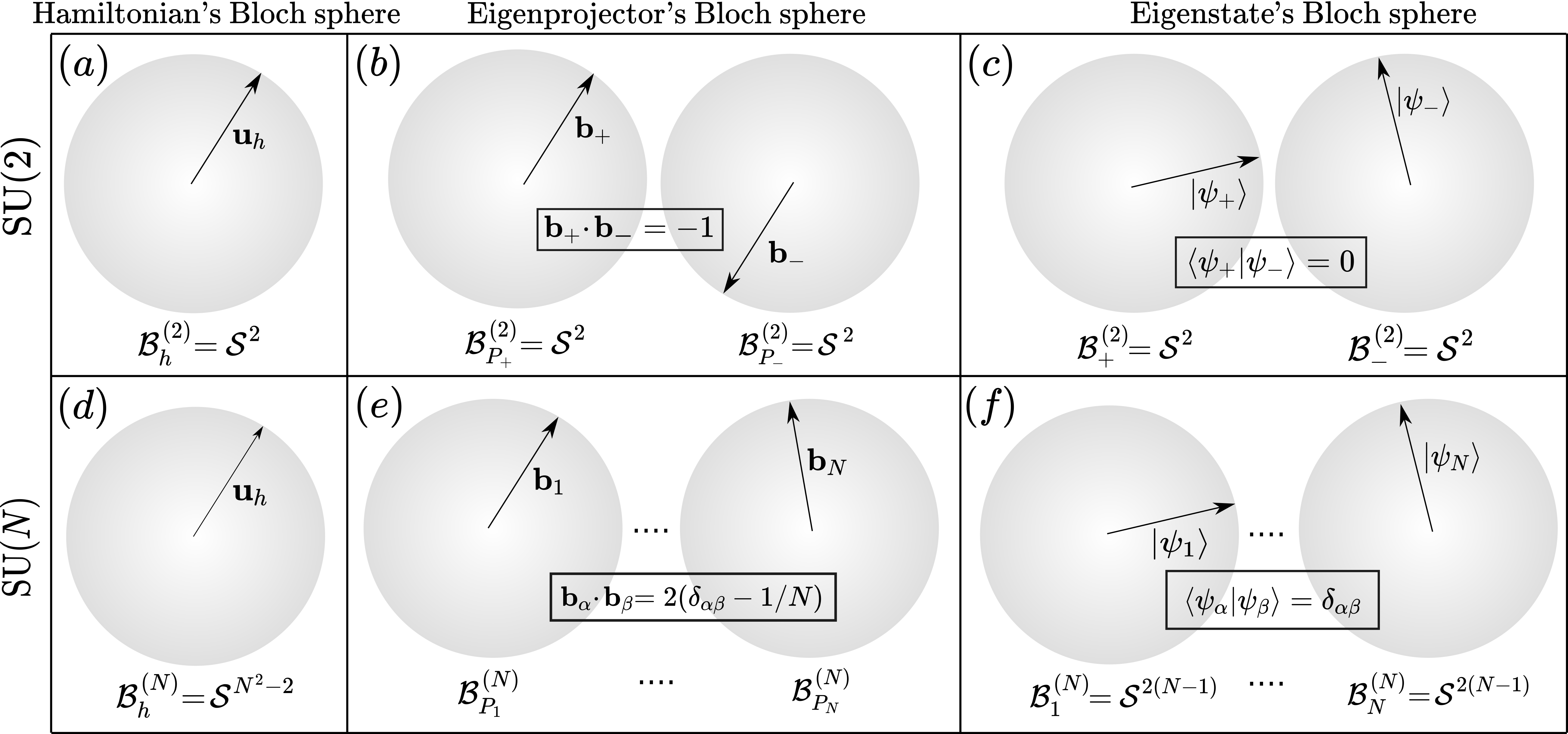}
	\caption{Schematic visualization of the different Bloch sphere's appearing in the treatment of SU($N$) Hamiltonians. The familiar SU($2$) case (Section \ref{SU2spheres}) is illustrated in the first row, while the second row treats the general case of arbitrary $N$ (Section \ref{SUNspheres}).  The Hamiltonian's Bloch sphere \smash{$\mathcal{B}_h^{(N)}$} is the relevant space for the mapping $\mathbf{u}_h(\mathbf{x})$, and corresponds to a proper unit sphere for all values of $N$. The eigenprojector's Bloch sphere \smash{$\mathcal{B}_{P\alpha}^{(N)}$} associated to the mapping $\mathbf{b}_\alpha(\mathbf{x})$ is only a proper sphere for $N=2$. In the $N>2$ case, where it is often called {\em generalized Bloch sphere}, it is of very complicated shape. The eigenstate's Bloch sphere \smash{$\mathcal{B}_\alpha^{(N)}$} associated to the mapping $\ket{\psi_\alpha(\mathbf{x})}$ is also a proper unit sphere for all $N$, but of different dimension than the Hamiltonian's Bloch sphere if $N>2$.}
	\label{fig:blochspherezoo}
\end{figure*}

\section{Short review of two-band systems}
\label{SU2}

\subsection{SU(2) Hamiltonian}

Consider a two-band system, described by a $2\times 2$ Hermitian matrix $H(\mathbf{x})$ with eigenvalues
$\{E_\alpha(\mathbf{x})|\,\alpha=\pm\}$ and (orthonormal) eigenstates $\{\ket{\psi_\alpha(\mathbf{x})}\}$. 
Without loss of generality, the trivial part of $H$ is set equal to zero, \ie $\Tr H=0$.
The Hamiltonian can be expanded in the Pauli matrices $\boldsymbol{\sigma}=(\sigma_x,\sigma_y,\sigma_z)$ as
\begin{equation}
	H(\mathbf{x})=\mathbf{h}(\mathbf{x})\cdot\boldsymbol{\sigma},
	\label{2def}
\end{equation}
with the \emph{Hamiltonian vector} $\mathbf{h}(\mathbf{x})=\Tr\{H\boldsymbol{\sigma}\}/2$. The eigenvalues of Eq. (\ref{2def}) are
\begin{equation}
	E_\alpha(\mathbf{x})=\alpha\sqrt{\Tr(H^2)/2}=\alpha|\mathbf{h}|.
	\label{en2}
\end{equation}

\subsection{Eigenprojectors and Bloch vectors}

The eigenprojectors $P_\alpha(\mathbf{x})\equiv\ket{\psi_\alpha(\mathbf{x})}\bra{\psi_\alpha(\mathbf{x})}$ of the Hamiltonian (\ref{2def}) can be expanded in the Pauli matrices as
\begin{equation} 
	\begin{aligned}
	P_\alpha(\mathbf{x})= \frac{1}{2}\left[1_2+\mathbf{b}_\alpha(\mathbf{x})\cdot\boldsymbol{\sigma}\right],
	\end{aligned}
	\label{defproj2}
\end{equation}	
which defines a \emph{Bloch vector} $\mathbf{b}_\alpha(\mathbf{x})$. It is related to the Hamiltonian vector as
\begin{equation}
	\mathbf{b}_\alpha(\mathbf{x})=\alpha\frac{\mathbf{h}}{|\mathbf{h}|}.
	\label{b2}
\end{equation}
Note the agreement with the important expression (\ref{P2b2}).

In contrast to the eigenstates, eigenprojectors and Bloch vectors are explicitly gauge-independent quantities and well-behaved in parameter space: a singularity can only appear at band touching points in parameter space. 

\subsection{Three different SU(2) Bloch spheres}
\label{SU2spheres}

Here, we briefly discuss three different notions of Bloch spheres encountered in the treatment of two-band systems. A precise distinction of these Bloch spheres is necessary to avoid confusion when introducing Bloch vectors for $N$-band Hamiltonians. 

The unit vector $\mathbf{u}_h(\mathbf{x})\equiv\mathbf{h}/|\mathbf{h}|$ associated to the Hamiltonian vector can be parametrized as $\mathbf{u}_h=(\sin\theta_h\cos\phi_h,\sin\theta_h\sin\phi_h,\cos\theta_h)$. The two parameters $(\theta_h(\mathbf{x}),\phi_h(\mathbf{x}))$ may be called Hamiltonian's angles, since $\cos \theta_h=h_z/|\mathbf{h}|$ and $\tan \phi_h=h_y/h_x$. The unit vector defines a map
\begin{equation}
u_h:\,X\rightarrow\smash{\mathcal{B}_h^{(2)}},\hspace{.5cm} \mathbf{x}\mapsto\mathbf{u}_h(\mathbf{x})
\end{equation}
 from the parameter space $X$ to the {\em Hamiltonian's Bloch sphere} \smash{$\mathcal{B}_h^{(2)}$}. It is simply a unit two-sphere, \smash{$\mathcal{B}_h^{(2)}=\mathcal{S}^2$}, see Fig. \ref{fig:blochspherezoo}(a).

The Bloch vector defines a map
\begin{equation}
\mathbf{b}_\alpha:\,X\rightarrow\smash{\mathcal{B}_{P_\alpha}^{(2)}},\hspace{.5cm} \mathbf{x}\mapsto\mathbf{b}_\alpha(\mathbf{x})
\end{equation}
from parameter space to a space that may be called the {\em eigenprojector's Bloch sphere} \smash{$\mathcal{B}_{P_\alpha}^{(2)}$}. Importantly, the peculiarity and simplicity of two-band systems, as compared to the more general $N$-band case, consists in the fact that the eigenprojector's Bloch sphere is a unit two-sphere, \smash{$\mathcal{B}_{P_\alpha}^{(2)}=\mathcal{S}^2$} [see Fig. \ref{fig:blochspherezoo}(b)], and that the Bloch vector is proportional to the Hamiltonian vector, cf. Eq. (\ref{b2}). 

The third type of Bloch sphere arises from the eigenstates. A two-band eigenstate can be parametrized as
\begin{equation}
		\ket{\psi_\alpha(\mathbf{x})}=e^{i\Gamma_\alpha}\begin{pmatrix}
			\cos\theta_\alpha\\
			\sin\theta_\alpha e^{i\phi_\alpha}
		\end{pmatrix},
	\label{2states}
\end{equation}
with a global phase $\Gamma_\alpha(\mathbf{x})$
and two eigenstate's angles $(\theta_\alpha(\mathbf{x}),\phi_\alpha(\mathbf{x}))$. These angles define a map 
\begin{equation}
\ket{\psi_\alpha}:\,X\rightarrow\smash{\mathcal{B}_\alpha^{(2)}},\hspace{.5cm} \mathbf{x}\mapsto\ket{\psi_\alpha(\mathbf{x})}
\end{equation}
from parameter space to an
\emph{eigenstate's Bloch sphere} \smash{$\mathcal{B}_\alpha^{(2)}$}. Since $(\theta_\alpha,\phi_\alpha)$ can be interpreted as spherical coordinates, \smash{$\mathcal{B}_\alpha^{(2)}$} is a unit two-sphere, too, see Fig. \ref{fig:blochspherezoo}(c). 
Note that the eigenstates (\ref{2states}) are explicitly gauge-dependent and potentially suffer from singularities, as discussed in Appendix \ref{App2states}. 

Although, in the two-band case, all three Bloch spheres happen to correspond to a unit two-sphere, it is important to keep in mind that they are {\em a priori distinct spaces}. This distinction is necessary to avoid considerable confusion in the SU($N>2$) case, where the three Bloch spheres are completely different from one another [see Fig. \ref{fig:blochspherezoo}(d)--(f) and Section \ref{SUNspheres}].

\subsection{Quantum geometric tensor}

%

In Eqs. (\ref{eq1})--(\ref{b23}), the explicit dimension and form of the matrices $g_\alpha$ and $\Omega_\alpha$ clearly depend on the number and type of parameters $x_i$ contained in the chosen set $\mathbf{x}$. Some examples for different interesting choices of $\mathbf{x}$ are given in Appendix \ref{Appparamchoice}.
If the set of parameters is kept completely general, the most convenient formulation of the QGT follows from Eq. (\ref{QGT23}):
\begin{equation}
\begin{aligned}
g_{\alpha,ij}(\mathbf{x})&=\frac{1}{4}\mathbf{b}_\alpha^i\cdot\mathbf{b}_\alpha^j\\
&=\frac{1}{4|\mathbf{h}|^2}\left[\mathbf{h}^i\cdot\mathbf{h}^j-\frac{\left(\mathbf{h}\cdot\mathbf{h}^i\right)\left(\mathbf{h}\cdot\mathbf{h}^j\right)}{|\mathbf{h}|^2}\right],\\
\Omega_{\alpha,ij}(\mathbf{x})&=-\frac{1}{2}\mathbf{b}_\alpha\cdot(\mathbf{b}_\alpha^i\times\mathbf{b}_\alpha^j)\\
&=-\frac{\alpha}{2|\mathbf{h}|^3}\mathbf{h}\cdot(\mathbf{h}^i\times\mathbf{h}^j),
\end{aligned}
\label{Q2}
\end{equation}
with the shorthand notation $\mathbf{m}^i\equiv\partial_i\mathbf{m}$. For obtaining the expressions in terms of Bloch vectors, we used Eq. (\ref{defproj2}) and the product identity
\begin{equation}
(\mathbf{m}\cdot\boldsymbol{\sigma})(\mathbf{n}\cdot\boldsymbol{\sigma})=\mathbf{m}\cdot\mathbf{n}\, 1_2+i(\mathbf{m}\times\mathbf{n})\cdot\boldsymbol{\sigma}.
\label{pauliprod}
\end{equation}
 The expressions in terms of $\mathbf{h}$ then follow from Eq. (\ref{P2b2}).

\section{Generalization to $N$-band systems: Part 1 -- Eigenprojectors and Bloch vectors}
\label{SUN}

We now systematically generalize the two-band discussion of the previous section to arbitrary $N$. More precisely, after introducing the generic $N$-band Hamiltonian (Section \ref{SUNham}), we will first generalize
Eq. (\ref{P2b2}), both for the eigenprojectors (Section \ref{projN}) and Bloch vectors (Section \ref{genbs}). The properties of the SU($N$) Bloch spheres are also briefly addressed (Section \ref{SUNspheres}). 

Once the eigenprojectors $P_\alpha(E_\alpha,H)$ and Bloch vectors $\mathbf{b}_\alpha(E_\alpha,\mathbf{h})$ are found, one can insert them into Eq. (\ref{QGT23}) or (\ref{b23}), to obtain the $N$-band generalization of the QGT (\ref{Q2}). This will be done in Section \ref{blochgeo}.

\subsection{SU($N$) Hamiltonian}
\label{SUNham}

Consider an $N$-band system, described by an $N\times N$ Hermitian matrix $H(\mathbf{x})$ with eigenvalues
$\{E_\alpha(\mathbf{x})|\,\alpha=1,...,N\}$ and (orthonormal) eigenstates $\{\ket{\psi_\alpha(\mathbf{x})}|\}$ \footnote{We assume the eigenvalues to be at most locally degenerate, such that single-band projectors and the notion of an Abelian QGT can be used.}. 

Again, the trivial part is set equal to zero without loss of generality, $\Tr H=0$.
The Hamiltonian can be expanded as
\begin{equation}
	H(\mathbf{x})=\mathbf{h}(\mathbf{x})\cdot\boldsymbol{\lambda},
	\label{genham0}
\end{equation}
with the Hamiltonian vector $\mathbf{h}(\mathbf{x})=\Tr\{H\boldsymbol{\lambda}\}/2$.
 Here, $\boldsymbol{\lambda}=(\lambda_1,...,\lambda_{N^2-1})$ 
is a vector composed of the $N^2-1$ elementary generator matrices of the SU$(N)$ Lie group. Together with the identity matrix $1_N$ they consitute a basis for the Lie algebra $\mathfrak{su}(N)$.

 Throughout this paper, we will choose the generators $\lambda_a$ as {\em (generalized) Gell-Mann} matrices \cite{Pfeifer_2003,Bertlmann_2008}. For the reader's convenience, the Gell-Mann matrices for $N=3$ and $N=4$ are listed in Appendix \ref{GellMann}.



\subsection{Eigenprojectors as a function of the Hamiltonian: $P_\alpha(E_\alpha,H)$}
\label{projN}

According to the Cayley-Hamilton theorem \cite{Cayley_1858,Horn_2013}, any function of the $N\times N$ Hamiltonian matrix $H$ can be written as a matrix polynomial in $H$, where the highest power is $N-1$. In particular, as demonstrated in detail in Appendix \ref{AppEle}, the eigenprojector $P_\alpha$ can be written as a matrix polynomial
\begin{equation}
\begin{aligned}
P_\alpha
&=\frac{\sum_{n=0}^{N-1}q_{N-1-n}(E_\alpha)H^n}{\sum_{n=0}^{N-1}q_{N-1-n}(E_\alpha)E_\alpha^n}\\
&=\frac{\sum_{n=0}^{N-1}q_{N-1-n}(E_\alpha)H^n}{\sum_{n=0}^{N-1}q_{N-1-n}(E_\alpha)C_n},
\end{aligned}
\label{Projectormain}
\end{equation}
where the polynomials
$q_n(z)\equiv\sum_{k=0}^{n}c_kz^{n-k}$
are closely related to the Hamiltonian's characteristic polynomial. The coefficients $c_k$ appearing here are listed in Table \ref{tabcn} for $k\le 5$, and depend solely on the \emph{(classical) Casimir invariants} \cite{Kusnezov_1995}
\begin{equation}
	C_n\equiv\Tr(H^n)=\sum_{\alpha=1}^{N}E_\alpha^n,
	\label{Cndef}
\end{equation}
where obviously $C_0=N$ and $C_1=0$.
\begin{table}[H]
	\centering
	\begin{tabular}{c||c|c|c|c|c|c}
		$k$  & 0&1&2&3&4&5\\
		\hline
		\hline
		$c_k$&1 &0 & $-\frac{C_2}{2}$ &$-\frac{C_3}{3}$  & $\frac{C_2^2}{8}-\frac{C_4}{4}$&  $\frac{C_2C_3}{6}-\frac{C_5}{5}$ 
	\end{tabular}
	\caption{Coefficients $c_k$ determining the polynomial $q_n(z)$.}
	\label{tabcn}
\end{table}
 
The utility of Eq. (\ref{Projectormain}) becomes immediately apparent if we write down the explicit eigenprojectors $P_\alpha(E_\alpha,H)$ for the cases $N=2$ to $N=5$:
\begin{widetext}
\begin{equation}
\begin{aligned} 
P_\alpha&=\frac{1}{2E_\alpha}(E_\alpha 1_2+H),\\ P_\alpha&=\frac{1}{3E_\alpha ^2-\frac{C_2 }{2}}\left[\left(E_\alpha^2-\frac{C_2 }{2}\right)1_3+E_\alpha H+H^2\right],\\ P_\alpha&=\frac{1}{4E_\alpha^3-C_2E_\alpha-\frac{C_3}{3}}
\left[\left(E_\alpha^3-\frac{C_2}{2}E_\alpha-\frac{C_3}{3}\right)1_4+\left(E_\alpha^2-\frac{C_2}{2}\right)H+E_\alpha H^2+H^3\right],\\
P_\alpha&=\frac{1}{5E_\alpha^4-\frac{3C_2}{2}E_\alpha^2-\frac{2C_3}{3}E_\alpha+\frac{C_2^2-2C_4}{8}}\left[\left(E_\alpha^4-\frac{C_2}{2}E_\alpha^2-\frac{C_3}{3}E_\alpha+\frac{C_2^2-2C_4}{8}\right)1_5\right.\\
&\hspace{6cm}\left.+\left(E_\alpha^3-\frac{C_2}{2}E_\alpha-\frac{C_3}{3}\right)H+\left(E_\alpha^2-\frac{C_2}{2}\right)H^2+E_\alpha H^3+H^4\right].
\end{aligned}
\label{Project34}
\end{equation}
\end{widetext}
This represents the $N$-band generalization of the two-band projectors (\ref{P2b2}).
Since the Hamiltonian is typically known for any given problem, the only ingredient required for explicitly computing the eigenprojector $P_\alpha$ is the corresponding eigenenergy $E_\alpha$. 

Beyond the computation of eigenprojectors, Eq. (\ref{Projectormain}) has two interesting applications. 
First, it can be used to rewrite any function $f(H)$ as a polynomial of order $N-1$ in the Hamiltonian, which corresponds to an alternative version of \emph{Sylvester's formula} \cite{Horn_2013}:
\begin{equation}
		f(H)=\sum_{\alpha=1}^Nf(E_\alpha)\frac{\sum_{n=0}^{N-1}q_{N-1-n}(E_\alpha)H^n}{\sum_{n=0}^{N-1}q_{N-1-n}(E_\alpha)E_\alpha^n}.
\end{equation}

Second, the eigenprojector is in some sense a more fundamental object than the eigenstate, namely Eq. (\ref{Projectormain}) may be employed for constructing energy eigenstates
\begin{equation}
	\begin{aligned}
		\ket {\psi_\alpha}=\frac{1}{\sqrt{\bra{\psi_g} P_\alpha(E_\alpha,H)  \ket {\psi_g}}} P_\alpha(E_\alpha,H)  \ket {\psi_g},
	\end{aligned}
	\label{psiNmain}
\end{equation}
by projecting onto a \emph{gauge freedom state} $\ket {\psi_g}$ that can be chosen arbitrarily (for more details, see Appendix \ref{ApppsifromP}). The eigenstates (\ref{psiNmain}) can then be used further
to compute matrix elements of observables, or the Berry connection and other quantities of interest.

\subsection{Bloch vectors as a function of the Hamiltonian vector: $\mathbf{b}_\alpha(E_\alpha,\mathbf{h})$}
\label{genbs}

The above key results (\ref{Projectormain}) \& (\ref{Project34}) on the eigenprojectors are more convenient for some practical purposes when reformulated in a vectorial language. More precisely, just like a Hamiltonian vector was introduced in Eq. (\ref{genham0}), one may define a {\em (generalized) Bloch vector} 
$\mathbf{b}_\alpha$ by expanding the eigenprojector as \cite{Hioe_1981}
\begin{equation}
P_\alpha(\mathbf{x})=\frac{1}{N}1_N+\frac{1}{2}\mathbf{b}_\alpha(\mathbf{x})\cdot\boldsymbol{\lambda}
\label{defproj}
\end{equation}
in the SU($N$) generators, where $\mathbf{b}_\alpha\equiv\Tr\{P_\alpha\boldsymbol{\lambda}\}$. Again, be reminded that throughout this paper we will always use the generators to be (generalized) Gell-Mann matrices. Importantly, however, the functional form of all SU($N$) vector identities discussed below is \emph{independent} of the particular choice of generator matrices.

Now, by inserting the expansions (\ref{genham0}) \& (\ref{defproj}) into Eq. (\ref{Projectormain}), one obtains the analog of the function $P_\alpha(E_\alpha,H)$ in the vectorial language, namely $\mathbf{b}_\alpha(E_\alpha,\mathbf{h})$, as explained  in Appendix \ref{Appblochder}. The usefulness of this procedure is most apparent when considering the explicit Bloch vector expressions resulting from it, provided here for $N=2$ to $N=5$:
\begin{widetext}
	\begin{equation}
		\begin{aligned}
			\mathbf{b}_\alpha&=\frac{1}{E_\alpha}\mathbf{h},\\ \mathbf{b}_\alpha&=\frac{2}{3E_\alpha^2-\frac{C_2}{2}}\left(E_\alpha\mathbf{h}+\mathbf{h}_\star\right),\\ \mathbf{b}_\alpha&=\frac{2}{4E_\alpha^3-C_2E_\alpha-\frac{C_3}{3}}\left[\left(E_\alpha^2-\frac{C_2}{4}\right)\mathbf{h}+E_\alpha\mathbf{h}_\star
			+\mathbf{h}_{\star\star}\right],\\
			\mathbf{b}_\alpha&=\frac{2}{5E_\alpha^4-\frac{3C_2}{2}E_\alpha^2-\frac{2C_3}{3}E_\alpha+\frac{C_2^2-2C_4}{8}}\left[\left(E_\alpha^3-\frac{3C_2}{10}E_\alpha-\frac{2C_3}{15}\right)\mathbf{h}+\left(E_\alpha^2-\frac{3C_2}{10}\right)\mathbf{h}_\star+E_\alpha\mathbf{h}_{\star\star}+\mathbf{h}_{\star\star\star}\right].
		\end{aligned}
		\label{balpha34}
	\end{equation}
\end{widetext}
This is the $N$-band generalization of the two-band Bloch vectors (\ref{P2b2}), and contains exactly the same information as Eq. (\ref{Project34}).

Two ingredients of Eq. (\ref{balpha34}) require some more explanation, namely the Casimir invariants $C_n$, and the "star product vectors" $\mathbf{h}_\star$, $\mathbf{h}_{\star\star}$, \etc~ Importantly, both quantities are completely determined by the vector $\mathbf{h}$. 

\subsubsection*{Casimir invariants from the Hamiltonian vector}

 Since the Casimir invariants are traces of powers of the Hamiltonian matrix, cf. Eq. (\ref{Cndef}), they can be directly expressed in terms of the Hamiltonian vector (cf. Appendix \ref{Appblochder}):
\begin{equation}
	\begin{aligned}
		C_2&=2|\mathbf{h}|^2, \\
		C_3&=2\,\mathbf{h}\cdot\mathbf{h}_\star,\\
		C_4&=4|\mathbf{h}|^4/N+2|\mathbf{h}_\star|^2.
	\end{aligned}
	\label{invariantsmain}
\end{equation}
It is convenient to combine Eq. (\ref{invariantsmain}) with Eq. (\ref{balpha34}), such that the Bloch vector for a given band $\alpha$ depends only on the single eigenvalue $E_\alpha$.

\subsubsection*{Star products of the Hamiltonian vector}

For given $N$, each Bloch vector (\ref{balpha34}) is a kind of "vector polynomial" of degree $N-1$ in $\mathbf{h}$, just like each eigenprojector is a matrix polynomial in $H$. In other words, the vectors $\mathbf{h}_\star$, $\mathbf{h}_{\star\star}$, \etc~-- though rather unfamiliar objects of dimension Energy$^2$, Energy$^3$, \etc~-- are uniquely determined by the Hamiltonian vector $\mathbf{h}$, as explained in the following.

The properties of the Lie algebra $\mathfrak{su}(N)$ underlying the $N$-band Hamiltonian (\ref{genham0}) are determined by the commutation and anticommutation relations \cite{Kaplan_1967}
\begin{equation}
	\begin{aligned}
		\comm{\lambda_a}{\lambda_b}&=2if_{abc}\lambda_c,\\
		\{\lambda_a,\lambda_b\}&=\frac{4}{N}\delta_{ab}1_N+2d_{abc}\lambda_c,
		\label{comms}
	\end{aligned}
\end{equation}
where repeated lower indices imply summation (Einstein convention). In particular, these relations define totally antisymmetric and totally symmetric \emph{structure constants} of $\mathfrak{su}(N)$, respectively:
\begin{equation}
	\begin{aligned}
f_{abc}&\equiv-\frac{i}{4}\Tr\left(\comm{\lambda_a}{\lambda_b}\lambda_c\right),\\
d_{abc}&\equiv\frac{1}{4}\Tr\left(\acomm{\lambda_a}{\lambda_b}\lambda_c\right).
\end{aligned}
\end{equation}
 These are a known set of real numbers once a matrix representation is chosen for the generators $\lambda_a$. Note that for $N=2$, where $\boldsymbol{\lambda}=\boldsymbol{\sigma}$, the $d_{abc}$ vanish identically and $f_{abc}=\epsilon_{abc}$, 
where $\epsilon_{abc}$ is the Levi-Civita tensor.

From the structure constants, one defines dot, star and cross products of SU($N$) vectors:
\begin{equation}
	\begin{aligned}
		\mathbf{m}\cdot\mathbf{n}&\equiv m_cn_c,\\
		(\mathbf{m}\star\mathbf{n})_a&\equiv d_{abc}m_bn_c,\\
		(\mathbf{m}\times\mathbf{n})_a&\equiv f_{abc}m_bn_c,\\
	\end{aligned}
	\label{prods}
\end{equation}
where $\mathbf{m}$ and $\mathbf{n}$ are real and $(N^2-1)$-dimensional. 
The star product is unfamiliar because it does not play any role in $N=2$ situations (where $d_{abc}=0$), but it is crucial for $N>2$ systems. Finally, we may introduce the notation for repeated star products of a vector with itself:
\begin{equation}
\begin{aligned}
	\mathbf{m}_\star^{(0)}&=\mathbf{m},\\
	\mathbf{m}_\star^{(1)}&=\mathbf{m}_\star\equiv\mathbf{m}\star\mathbf{m},\\
	\mathbf{m}_\star^{(2)}&=\mathbf{m}_{\star\star}\equiv\mathbf{m}\star(\mathbf{m}\star\mathbf{m}),\\
	\mathbf{m}_\star^{(k+1)}&=\mathbf{m}_{\star\star...}\equiv\mathbf{m}\star\mathbf{m}_\star^{(k)}.
\end{aligned}
\label{stardef}
\end{equation}
The vectors $\mathbf{h}_\star$, $\mathbf{h}_{\star\star}$, \etc~are thus simply star products of the Hamiltonian vector with itself.

\subsection{Three different SU($N$) Bloch spheres}
\label{SUNspheres}

The Hamiltonian's, eigenprojector's and eigenstate's Bloch spheres introduced in Section \ref{SU2spheres} for the two-band case are here briefly discussed for higher $N$.

The Hamiltonian vector $\mathbf{h}$ has $N^2-1$ real-valued components,
such that the unit vector $\mathbf{u}_h\equiv\mathbf{h}/|\mathbf{h}|$ can be parametrized by $N^2-2$ Hamiltonian's angles. In other words, $\mathbf{u}_h(\mathbf{x})$ defines a map 
\begin{equation}
	u_h:\,X\rightarrow\smash{\mathcal{B}_h^{(N)}},\hspace{.5cm} \mathbf{x}\mapsto\mathbf{u}_h(\mathbf{x})
\end{equation}
from the parameter space to a {\em Hamiltonian's Bloch sphere} \smash{$\mathcal{B}_h^{(N)}=\mathcal{S}^{N^2-2}$}, where $\mathcal{S}^{N^2-2}$ is the unit $(N^2-2)$-sphere, as depicted in Fig. \ref{fig:blochspherezoo}(d). 

Naively, since $\mathbf{b}_\alpha$ also has $N^2-1$ real-valued components, the map $\mathbf{b}_\alpha(\mathbf{x})$ (for given $\alpha$) would similarly seem to define an $(N^2-2)$-sphere, but this is prevented by constraints on the Bloch vectors.
In particular, the usual orthogonality relation $P_\alpha P_\beta=\delta_{\alpha\beta}P_\alpha$ and completeness relation \smash{$\sum_\alpha P_\alpha=1_N$} translate to the Bloch vector picture as
\begin{equation}
	\begin{aligned}
		\mathbf{b}_\alpha\cdot\mathbf{b}_\beta&=2\left(\delta_{\alpha\beta}-\frac{1}{N}\right),\\
		\mathbf{b}_\alpha\star\mathbf{b}_\beta&=\left(\delta_{\alpha\beta}-\frac{2}{N}\right)(\mathbf{b}_\alpha+\mathbf{b}_\beta),\\
		\mathbf{b}_\alpha\times\mathbf{b}_\beta&=0,\\
		\sum_\alpha\mathbf{b}_\alpha&=0.
		\label{unitbloch}
	\end{aligned}
\end{equation}
As a consequence, the vector $\mathbf{b}_\alpha(\mathbf{x})$ defines a map
\begin{equation}
	\mathbf{b}_\alpha:\,X\rightarrow\smash{\mathcal{B}_{P_\alpha}^{(N)}},\hspace{.5cm} \mathbf{x}\mapsto\mathbf{b}_\alpha(\mathbf{x})
\end{equation}
 from parameter space not to an $(N^2-2)$-sphere but to a $2(N-1)$-dimensional subset thereof, which may be called the {\em eigenprojector's Bloch sphere} \smash{$\mathcal{B}_{P_\alpha}^{(N)}$}, or simply the {\em generalized Bloch sphere}, see Fig. \ref{fig:blochspherezoo}(e). In contrast to the $N=2$ case, the Bloch vector is no longer parallel to the Hamiltonian vector for $N>2$ [cf. Eq. (\ref{balpha34})], such that
 $\mathbf{u}_h(\mathbf{x})$ and $\mathbf{b}_\alpha(\mathbf{x})$ are completely distinct maps. An understanding of the true geometrical structure of the generalized Bloch sphere is not at all easy to acquire. Many efforts have been undertaken to figure out its properties for $N>2$, which is a surprisingly nontrivial issue, see Refs. \cite{Harriman_1978,Jakobczyk_2001,Kimura_2003,Zyczkowski_2003,Byrd_2003,Kimura_2005,Mendas_2006,Goyal_2016} and references therein. For the interested reader, we outline the main results in Appendix \ref{geosphere}. 

The third type of Bloch sphere arises from the eigenstates. Each eigenstate (\ref{psiNmain}) of an $N$-band system can be minimally encoded by a global phase $\Gamma_\alpha(\mathbf{x})$ and 
$N-1$ pairs of eigenstate's angles $(\theta_\alpha ^i(\mathbf{x}),\phi_\alpha ^i(\mathbf{x}))$ ($i=1,...,N-1$). 
These angles define a map 
\begin{equation}
	\ket{\psi_\alpha}:\,X\rightarrow\smash{\mathcal{B}_\alpha^{(N)}},\hspace{.5cm} \mathbf{x}\mapsto\ket{\psi_\alpha(\mathbf{x})}
\end{equation}
from the parameter space to a $2(N-1)$-dimensional \emph{eigenstate's Bloch sphere} \smash{$\mathcal{B}_\alpha^{(N)}$}, 
which is depicted, for comparison with the $N=2$ case, in Fig. \ref{fig:blochspherezoo}(f). 

In summary, the Hamiltonian's, the eigenprojector's and the eigenstate's Bloch sphere are all isomorphic spaces for $N=2$, but all different spaces for $N>2$.

\section{Generalization to $N$-band systems: Part 2 -- Quantum geometric tensor}
\label{blochgeo}

The QGT in $N$-band systems is usually defined in terms of $N$-component eigenstates $\ket{\psi_\alpha}$. The standard eigenstate-based QGT formula is given by Eq. (\ref{QG0}). (For more details on its origin, see Appendix \ref{AppQGT}.) A more popular eigenstate-based QGT expression, which is easily found by treating $H(\mathbf{x}+d\mathbf{x})-H(\mathbf{x})$ as a formal perturbation to $\ket{\psi_\alpha(\mathbf{x})}$, reads
\begin{equation}
	T_{\alpha,ij}=\sum_{\beta\neq\alpha}\frac{\bra{\psi_\alpha }\partial_iH \ket{\psi_\beta}\bra{\psi_\beta}\partial_jH \ket{\psi_\alpha}}{(E_\alpha-E_\beta)^2}.
	\label{nextQGT}
\end{equation}
The advantage of Eq. (\ref{nextQGT}), as compared to Eq. (\ref{QG0}), is that it involves the parametric velocity operators $\partial_iH$, thus avoiding derivatives of eigenstates. Therefore, it is the most common formula used for computing the QGT. However, this in general still requires \emph{numerical construction} of the eigenstates.

The main goal here is to develop a deeper \emph{analytical understanding} of the geometric tensors and their direct relation to the Hamiltonian. This can be achieved by avoiding eigenstates and drawing on the eigenprojector-based formula (\ref{QGT23}). (For the relation between Eq. (\ref{QG0}) and Eq. (\ref{QGT23}), see Appendix \ref{AppQGT}.)

First, we will reformulate Eq. (\ref{QGT23}) in terms of generalized Bloch vectors (Section \ref{altpics}). Then, inserting Eq. (\ref{balpha34}) into the result, we will be able to write the QGT in terms of only the Hamiltonian vector $\mathbf{h}$ and the eigenenergy $E_\alpha$  (Section \ref{explsec}).

\subsection{QGT in terms of Bloch vectors}
\label{altpics}

In order to obtain a Bloch vector picture of the geometric tensors, one can simply substitute Eq. (\ref{defproj}) into Eq. (\ref{QGT23}). Using the SU($N$) product identity (cf. Appendix \ref{Appblochder})
\begin{equation}
	(\mathbf{m}\cdot\boldsymbol{\lambda})(\mathbf{n}\cdot\boldsymbol{\lambda})=\frac{2}{N}\mathbf{m}\cdot\mathbf{n}\,
	1_N+(\mathbf{m}\star\mathbf{n}+i\,\mathbf{m}\times\mathbf{n})\cdot\boldsymbol{\lambda}
	\label{SUNid}
\end{equation}
 and the star product (\ref{unitbloch}) of Bloch vectors, one finds
\begin{equation}
\begin{aligned}
g_{\alpha,ij}&=\frac{1}{4}\mathbf{b}_\alpha^i\cdot\mathbf{b}_\alpha^j,\\
\Omega_{\alpha,ij}&=-\frac{1}{2}\mathbf{b}_\alpha\cdot\left(\mathbf{b}_\alpha^i\times\mathbf{b}_\alpha^j\right),
\label{geom}
\end{aligned}
\end{equation}
for \emph{arbitrary} values of $N$ (again with the shorthand notation $\mathbf{m}^i\equiv\partial_i\mathbf{m}$). This is the same as Eq. (\ref{b23}) stated in the introduction.

Symbolically, Eq. (\ref{geom}) is of the exactly same form as in the $N=2$ case, cf. Eq. (\ref{Q2}). However, be aware that for $N>2$ the dimension of the Bloch vectors increases, and the cross product has to be interpreted in the SU($N$) sense, cf. Eq. (\ref{prods}). 
For the SU($3$) case, an equivalent way of writing Berry curvature and quantum metric was already encountered in Refs.  \cite{Barnett_2012, Lee_2015} and \cite{Bauer_2016}, respectively; for the SU($N$) case, Pozo and de Juan recently found a similar formula in terms of what they call \emph{1-generators} \cite{Pozo_2020}. 

While Eq. (\ref{geom}) is conceptually important, a certain disadvantage consists in the presence of explicit parametric derivatives of the Bloch vector; this is potentially complicated, since cumbersome expressions follow from applying the product rule to Eq. (\ref{balpha34}). 
An alternative form of writing the QGT that does not involve such explicit derivatives of the Bloch vectors is provided in Appendix \ref{Appblochalt}.


\subsection{QGT from the Hamiltonian and its eigenvalues}
\label{explsec}

Having established the general expressions (\ref{geom}) for arbitrary values of $N$, 
the final step towards obtaining the QGT in terms of only the Hamiltonian vector $\mathbf{h}$ and
the eigenenergy $E_\alpha$ consists in inserting the explicit formula (\ref{balpha34}).  

For the Berry curvature, we obtain a closed form expression for arbitrary $N$, which is illustrated here for $N=3,4,5$ for comparison with the simple $N=2$ expression (\ref{Q2}); the discussion for arbitrary $N$ is given in Appendix \ref{AppBerry}. 

In the $N=3$ case,
the Berry curvature tensor is given by
\begin{equation}
\Omega_{\alpha,ij}=-\frac{4(E_\alpha\mathbf{h}+\mathbf{h}_\star)}{\left(3E_\alpha^2-|\mathbf{h}|^2\right)^3}
\cdot\left[\left(E_\alpha\mathbf{h}^i+\mathbf{h}_\star^i\right)\times\left(E_\alpha\mathbf{h}^j+\mathbf{h}_\star^j\right)\right],
\label{enbar}
\end{equation}
with $\mathbf{h}_\star^i\equiv\partial_i\mathbf{h}_\star$. 
This expression is equivalent to the result found by Barnett \emph{et al.} \cite{Barnett_2012} if
 one inserts a closed-form energy parametrization in terms of trigonometric functions, see for example Ref. \cite{Rosen_1971}. 
Note that Eq. (\ref{enbar}) can be expressed in a more compact form (cf. Appendix \ref{AppBerry}), which allows to explicitly verify the sum rule $\sum_\alpha\Omega_{\alpha,ij}=0$. 

For arbitrary $N=4$ systems, we have
\begin{widetext}
\begin{equation}
\Omega_{\alpha,ij}=-\frac{4(Q_\alpha\mathbf{h}+E_\alpha\mathbf{h}_\star+\mathbf{h}_{\star\star})}
{(4E_\alpha Q_\alpha-\frac{2}{3}\mathbf{h}\cdot\mathbf{h}_\star)^3}\cdot\left[\left(Q_\alpha\mathbf{h}^i+E_\alpha\mathbf{h}_\star^i+\mathbf{h}_{\star\star}^i\right)
\times\left(Q_\alpha\mathbf{h}^j+E_\alpha\mathbf{h}_\star^j+\mathbf{h}_{\star\star}^j\right)\right],
\label{om4}
\end{equation}
with $Q_\alpha\equiv E_\alpha^2-|\mathbf{h}|^2/2$.
Similarly, for arbitrary $N=5$ systems, the Berry curvature is given by
\begin{equation}
	\begin{aligned}
		\Omega_{\alpha,ij}&=-\frac{4(R_\alpha\mathbf{h}+\tilde{R}_\alpha\mathbf{h}_\star+E_\alpha\mathbf{h}_{\star\star}+\mathbf{h}_{\star\star\star})}{\left(5E_\alpha R_\alpha+\frac{3}{10}|\mathbf{h}|^4-\frac{1}{2}|\mathbf{h}_\star|^2\right)^3}\\
		&\hspace{1cm}\cdot\left[(R_\alpha\mathbf{h}^i+\tilde{R}_\alpha\mathbf{h}_\star^i+E_\alpha\mathbf{h}_{\star\star}^i+\mathbf{h}_{\star\star\star}^i)\times(R_\alpha\mathbf{h}^j+\tilde{R}_\alpha\mathbf{h}_\star^j+E_\alpha\mathbf{h}_{\star\star}^j+\mathbf{h}_{\star\star\star}^j)\right],
	\end{aligned}
	\label{om5}
\end{equation}
\end{widetext}
where $R_\alpha\equiv E_\alpha\tilde{R}_\alpha-\frac{4}{15}\mathbf{h}\cdot\mathbf{h}_\star$ and $\tilde{R}_\alpha\equiv E_\alpha^2-\frac{3}{5}|\mathbf{h}|^2$. In the same way, one may obtain expressions for $N>5$.

For the quantum metric, writing down explicit formulas in terms of $\mathbf{h}$ and $E_\alpha$ proves too cumbersome for $N>2$, due to the absence of appropriate orthogonality relations. The quantum metric can however be straightforwardly computed by a two-step procedure: first calculate the Bloch vectors (\ref{balpha34}), then substitute the result into Eq. (\ref{geom}).

\section{Examples: Multifold fermions}
\label{examples}

In this section, the utility of the formalism developed above is demonstrated by applying it to concrete Hamiltonians of interest. In particular, we will consider $N=3$ and $N=4$ models in the class of \emph{multifold (Dirac) fermions} \cite{Flicker_2018}, \ie low-energy models $H(\mathbf{q})$ with a linear spectrum and an $N$-fold degeneracy at $\mathbf{q}=0$. Here, $\mathbf{x}=\mathbf{q}$ should be understood as some generic three-dimensional quasi-momentum vector, which may take various physical meanings \cite{Zak_1989,Dalibard_2011,Qi_2011,Armitage_2018,Cooper_2019,Ozawa_2019,ZhangD_2019}.

\subsection{Pseudospin fermions}

As a first simple class of models that lends itself to an analysis within the above language, consider (pseudo)spin-$S$ multifold fermion models, where $S=(N-1)/2$. Such models involve matrices $S_{x,y,z}$ that satisfy a spin algebra
\begin{equation}
	\comm{S_i}{S_j}=i\epsilon_{ijk}S_k,\hspace{.5cm}\text{for any triple $(i,j,k)$},
	\label{spinalg}
\end{equation}
as well as $\mathbf{S}^2=S(S+1) 1_N$. The corresponding Hamiltonian can be written as 
\begin{equation}
	\begin{aligned}
		H(\mathbf{q})&=\mathbf{q\cdot\mathbf{S}}=q_xS_x+q_yS_y+q_zS_z.
	\end{aligned}
	\label{berryspin}
\end{equation}
It may be checked that this Hamiltonian has a (global) charge conjugation symmetry $\mathcal{C}^\dagger H^*(\mathbf{q})\mathcal{C}=-H(\mathbf{q})$ \cite{Ludwig_2015}, with $\mathcal{C}=e^{i \pi S_y}$ and $\mathcal{CC}^*=(-1)^{2S}1_{2S+1}$. The spectrum is symmetric about zero energy, as well as isotropic and linear in the quasi-momentum:
\begin{equation}
E_m=m|\mathbf{q}|,
\label{spinspec}
\end{equation}
where $m=-S,...,S$. 

In the conventional eigenstate picture, the QGT of a pseudospin-$S$ fermion can be computed as follows. One may construct the $\mathbf{q}$-dependent
spin eigenstates $|S,m, \mathbf{q}\rangle$, where $m=-S,...,+S$, as rotated eigenstates of the form
$|S,m, \mathbf{q}\rangle =e^{i\varphi_\mathbf{q}S_z}e^{i\theta_\mathbf{q}S_y}|S,m\rangle$, with
$S_z|S,m\rangle=m|S,m\rangle$, $\cos \theta_{\mathbf{q}}=q_z/|\mathbf{q}|$, and $\tan \varphi_{\mathbf{q}}=q_y/q_x$.
One may then use Eq. (\ref{nextQGT}) to obtain the QGT $T_{m,ij}(\mathbf{q})$. In particular, defining the three-component pseudo-vector 
$\boldsymbol{\Omega}_m\equiv(\Omega_{m,yz},\Omega_{m,zx},\Omega_{m,xy})$, the Berry curvature takes the form of a topological monopole \cite{Berry_1984}:
\begin{equation}
	\boldsymbol{\Omega}_m(\mathbf{q})=-m\frac{\mathbf{q}}{|\mathbf{q}|^3}. 
	\label{omberry}
\end{equation}
This monopole carries a topological charge measured by the first Chern number, $\mathcal{C}_m=-2m$.
As is well known, $\mathcal{C}_m$ is odd (even) for half-integer (integer) spin.

\begin{figure}
	\centering
	\includegraphics[width=\columnwidth]{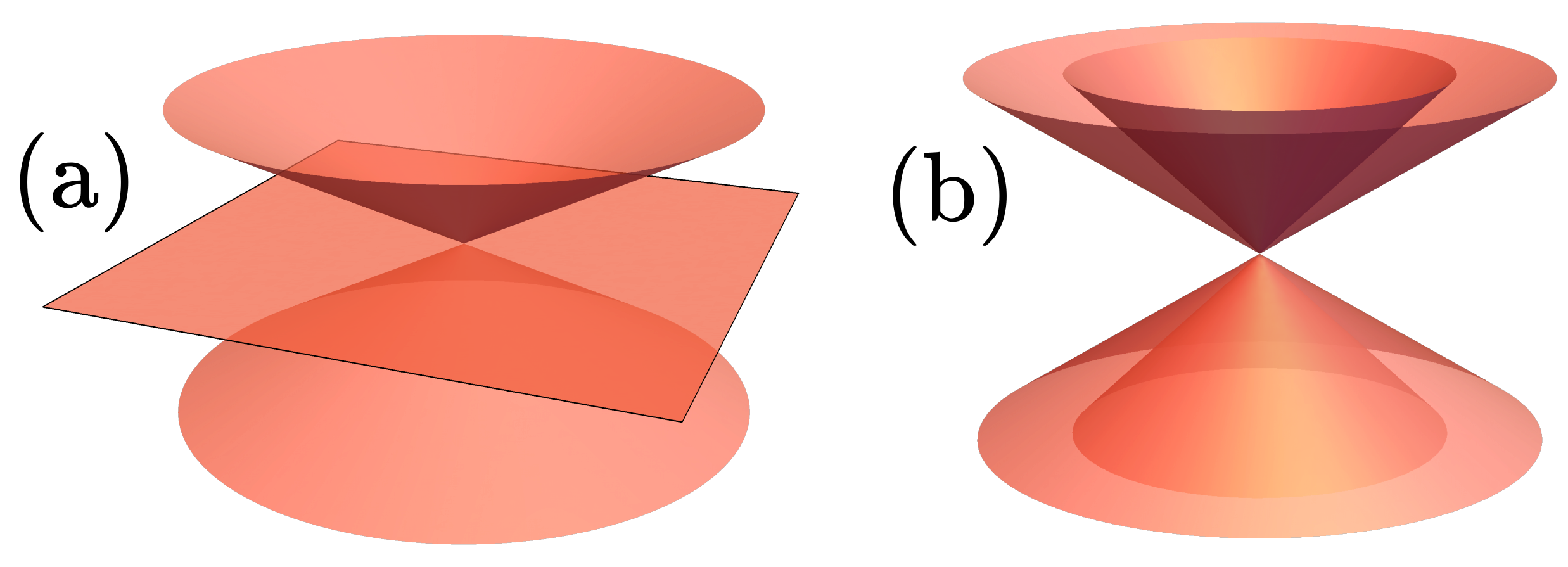}
	\caption{(a) Energy spectrum of a pseudospin-1 fermion and of the model (\ref{3mod}). (b) Energy spectrum of a pseudospin-3/2 fermion and of the model (\ref{HAandB}).}
	\label{fig:lowen}
\end{figure}

We now restrict ourselves to the three- and four-band case for simplicity, \ie to pseudospin-1 and pseudospin-3/2 fermions, respectively. The respective band structures are shown in Fig. \ref{fig:lowen}(a)\&(b).
Our goal is to obtain the eigenprojectors and Bloch vectors, which will allow us to compute the QGT without constructing spin eigenstates. In particular, we will recover the result (\ref{omberry}) and additionally compute the quantum metric.  We now separate the treatment for spin-1 and spin-3/2.

\subsubsection*{Pseudospin-1 fermions}

The spin Hamiltonian (\ref{berryspin}) can be easily rewritten in the general form of Eq. (\ref{genham0}) by using the relation between spin matrices and Gell-Mann matrices.
In the standard representation, the spin-1 matrices are related to the $N=3$ Gell-Mann matrices by Eq. (\ref{standardrep}). Thus, the Hamiltonian vector reads
\begin{equation}
		\mathbf{h}(\mathbf{q})=\frac{1}{2}(\sqrt{2}q_x,\sqrt{2}q_y,q_z,0,0,\sqrt{2}q_x,\sqrt{2}q_y,\sqrt{3}q_z).
	\label{hhstar}
\end{equation} 
Upon using the spectrum (\ref{spinspec}) and $C_2=2|\mathbf{q}|^2$ [cf. Eq. (\ref{invariantsmain})], the eigenprojectors for a pseudospin-1 fermion are obtained from Eq. (\ref{Project34}) as
\begin{equation}
	\begin{aligned}
		P_m(\mathbf{q})&=\frac{1}{3m^2-1}\left[(m^2-1)1_3+\frac{m}{|\mathbf{q}|}(\mathbf{h}\cdot\boldsymbol{\lambda})\right.\\
		&\left.\hspace{2cm}+\frac{1}{|\mathbf{q}|^2}(\mathbf{h}\cdot\boldsymbol{\lambda})^2\right].
	\end{aligned}
	\label{Pspin1}
\end{equation} 
It is readily verified that $\Tr(P_m)=1$ and $\sum_m P_m=1_3$ as expected. Similarly, the Bloch vectors follow from Eq. (\ref{balpha34}) as
\begin{equation}
	\begin{aligned}
		\mathbf{b}_m(\mathbf{q})&=\frac{2}{3m^2-1}\left(m\frac{\mathbf{h}}{|\mathbf{q}|}+\frac{\mathbf{h}_\star}{|\mathbf{q}|^2}\right),
	\end{aligned}
	\label{bspin1}
\end{equation} 
where
\begin{equation}
	\begin{aligned}
		\mathbf{h}_\star(\mathbf{q})&=\left(\frac{q_xq_z}{\sqrt{2}},\frac{q_yq_z}{\sqrt{2}},\frac{3q_z^2-|\mathbf{q}|^2}{4},\frac{q_x^2-q_y^2}{2},\right.\\
		&\hspace{1cm}\left.q_xq_y,-\frac{q_xq_z}{\sqrt{2}},-\frac{q_yq_z}{\sqrt{2}},\frac{|\mathbf{q}|^2-3q_z^2}{4\sqrt{3}}\right).
	\end{aligned}
\end{equation} 
The Berry curvature may now be obtained from Eq. (\ref{geom}) [or more directly from Eq. (\ref{enbar})], and one recovers Eq. (\ref{omberry}) as expected. Similarly, the quantum metric follows from Eq. (\ref{geom}) as
\begin{equation}
	g_{m,ij}(\mathbf{q})=\frac{2-m^2}{2|\mathbf{q}|^2}\left(\delta_{ij}-\frac{q_i q_j}{|\mathbf{q}|^2}\right),
	\label{metricS1}
\end{equation}
in agreement with Ref. \cite{Lin_2021}.

\subsubsection*{Pseudospin-3/2 fermions}

In the standard representation, the spin-3/2 matrices are related to the $N=4$ Gell-Mann matrices by Eq. (\ref{spin4}). The Hamiltonian vector thus reads
\begin{equation}
	\begin{aligned}
		\mathbf{h}(\mathbf{q})&=\frac{1}{2}(\sqrt{3}q_x,\sqrt{3}q_y,q_z,0,0,2 q_x,2 q_y,\sqrt{3}q_z,\\
		&\hspace{.8cm}0,0,0,0,\sqrt{3}q_x,\sqrt{3}q_y,\sqrt{6}q_z).
	\end{aligned}
	\label{h3/2}
\end{equation}
The star product vectors $\mathbf{h}_\star$ and $\mathbf{h}_{\star\star}$ can be computed from Eq. (\ref{h3/2}) using Eq. (\ref{stardef}).

Upon using the spectrum (\ref{spinspec}), as well as $C_2=5|\mathbf{q}|^2$ and $C_3=0$, the eigenprojectors for a spin-3/2 fermion are obtained as
\begin{equation}
	\begin{aligned}
		P_{m}(\mathbf{q})&=\frac{1}{4m\left(m^2-\frac{5}{4}\right)}\left[m \left(m^2-\frac{5}{2}\right)1_4\right.\\
		&\left.+\frac{m^2-\frac{5}{2}}{|\mathbf{q}|}(\mathbf{h}\cdot\boldsymbol{\lambda})+\frac{m}{|\mathbf{q}|^2}(\mathbf{h}\cdot\boldsymbol{\lambda})^2+\frac{1}{|\mathbf{q}|^3}(\mathbf{h}\cdot\boldsymbol{\lambda})^3\right].
	\end{aligned}
\end{equation}
Similarly, the Bloch vectors are given by
\begin{equation}
	\begin{aligned}
		\mathbf{b}_{m}(\mathbf{q})&=\frac{1}{2m \left(m^2-\frac{5}{4}\right)}\left[\left(m^2-\frac{5}{4}\right)\frac{\mathbf{h}}{|\mathbf{q}|}\right.\\
		&\left.\hspace{2.7cm}+m\frac{\mathbf{h}_\star}{|\mathbf{q}|^2}+
		\frac{\mathbf{h}_{\star\star}}{|\mathbf{q}|^3}\right].
		\label{bspin2}
	\end{aligned}
\end{equation}
 Importantly, be aware that the coefficient of the linear term (in $H$) of the eigenprojector is in general not the same as the coefficient of the linear term (in $\mathbf{h}$) of the Bloch vector, and similarly for the higher-order terms. This is due to the fact that \smash{$H^{n+1}\neq\mathbf{h}_\star^{(n)}\cdot\boldsymbol{\lambda}$} for $n\geq0$.

The Berry curvature follows from Eq. (\ref{geom}) [or more directly from Eq. (\ref{om4})], and again takes the expected form (\ref{omberry}). Similarly, the quantum metric reads
\begin{equation}
	g_{m,ij}(\mathbf{q})=\frac{15/4-m^2}{2|\mathbf{q}|^2}\left(\delta_{ij}-\frac{q_i q_j}{|\mathbf{q}|^2}\right),
	\label{metricS}
\end{equation}
again in agreement with Ref. \cite{Lin_2021}.

\subsection{Beyond pseudospin fermions}

Let us now discuss two examples that go beyond the simple pseudospin models. More precisely, we present a three-band (four-band) multifold fermion model with \emph{exactly the same} energy spectrum as a spin-1 (spin-3/2), cf. Fig. \ref{fig:lowen}, but with completely different symmetries and quantum geometric properties. It is important to realize that the analytical expressions presented below are easily obtained using our approach, while establishing them from an energy eigenstate approach would be rather cumbersome. 

\subsubsection*{Three-band example}

Consider the Hamiltonian
\begin{equation}
	\begin{aligned}
		H(\mathbf{q})&=\begin{pmatrix}
			0 & \frac{1}{\sqrt{2}} (q_x-iq_y) & -i q_z\\
			... & 0 & \frac{1}{\sqrt{2}}(q_x+iq_y)\\
			...& ...&0
		\end{pmatrix},
	\end{aligned}
	\label{3mod}
\end{equation}
where the lower left matrix elements are obtained by complex conjugating the upper right ones. 
This model has a global chiral symmetry $\mathcal{S}^\dagger H(\mathbf{q})\mathcal{S}=-H(\mathbf{q})$ \cite{Ludwig_2015} with $\mathcal{S}=\text{antidiag}(1,-1,1)$. The energy spectrum reads
\begin{equation}
	E_\alpha(\mathbf{q})=\alpha|\mathbf{q}|,\hspace{.5cm} \alpha=0,\pm1,
	\label{eps3}
\end{equation} 
which is the same as for a pseudospin-1 model.

In the language of Eq. (\ref{genham0}), the model (\ref{3mod}) is characterized by a Hamiltonian vector
\begin{equation}
\mathbf{h}(\mathbf{q})=\frac{1}{\sqrt{2}}(q_x,q_y,0,0,\sqrt{2}q_z,q_x,q_y,0)
\end{equation}
and its star product
\begin{equation}
	\begin{aligned}
		\mathbf{h}_\star(\mathbf{q})&=\left(\frac{q_yq_z}{\sqrt{2}},\frac{q_xq_z}{\sqrt{2}},\frac{3q_z^2-|\mathbf{q}|^2}{4},\frac{q_x^2-q_y^2}{2},\right.\\
		&\hspace{.8cm}\left.q_xq_y,\frac{q_yq_z}{\sqrt{2}},\frac{q_xq_z}{\sqrt{2}},\frac{|\mathbf{q}|^2-3q_z^2}{4\sqrt{3}}\right).
	\end{aligned}
	\label{hhstar2}
\end{equation} 
The Bloch vectors follow from Eq. (\ref{balpha34}) as
\begin{equation}
	\begin{aligned}
		\mathbf{b}_\alpha(\mathbf{q})&=\frac{2}{3\alpha^2-1}\left(\alpha\frac{\mathbf{h}}{|\mathbf{q}|}+\frac{\mathbf{h}_\star}{|\mathbf{q}|^2}\right).
	\end{aligned}
\label{ba}
\end{equation}
Notably, they are formally completely equivalent to the spin-1 case, but with slightly different vectors $\mathbf{h}$ and $\mathbf{h}_\star$.

The QGT for all bands $\alpha=0,\pm1$ can be summarized in compact form as
\begin{equation}
	\begin{aligned}
		\boldsymbol{\Omega}_\alpha(\mathbf{q})&=(2-3\alpha^2)(\mathbf{q}\cdot\mathbf{e}_z)\frac{\mathbf{q}}{|\mathbf{q}|^4},\\
		g_{\alpha,ij}(\mathbf{q})&=\frac{2-\alpha^2}{2|\mathbf{q}|^2}\left[\delta_{ij}-\frac{q_iq_j}{|\mathbf{q}|^2}\right.\\
		&\left.\hspace{1.7cm}-\frac{2-\alpha^2}{2|\mathbf{q}|^2}(\mathbf{q}\times\mathbf{e}_z)_i(\mathbf{q}\times\mathbf{e}_z)_j\right],
		\label{tunab}
	\end{aligned}
\end{equation}
where we again introduced a Berry curvature pseudovector $\boldsymbol{\Omega}_\alpha$ and where $\mathbf{e}_z=(0,0,1)$.
There are several notable features of Eq. (\ref{tunab}). First, the Berry curvature is completely different from its spin-1 analog (\ref{omberry}) and yields first Chern numbers $\mathcal{C}_\alpha=0$. Second, the quantum metric contains a part that is exactly like for a spin-1 [cf. Eq. (\ref{metricS1})], plus an additional term. Third, the QGT is anisotropic despite an isotropic band structure.

\subsubsection*{Four-band example}

Consider now the Hamiltonian
\begin{equation}
			H(\mathbf{q})=\begin{pmatrix}
				0 & q_y+\frac{q_z}{2} & -i \left(q_x+\frac{q_y}{2}\right) & \frac{q_x}{2}-q_z\\
				... & 0 & \frac{q_x}{2}+q_z & i \left(q_x-\frac{q_y}{2}\right)\\
				...&...&0&q_y-\frac{q_z}{2}\\
				...& ...&...&0
			\end{pmatrix}.
	\label{HAandB}
\end{equation}
This model has a global charge conjugation symmetry with $\mathcal{C}=\text{diag}(1,-1,1,-1)$, but no chiral symmetry, and a band structure
\begin{equation}
E_{\alpha_1\alpha_2}(\mathbf{q})=\alpha_1\left(1+\frac{\alpha_2}{2}\right)|\mathbf{q}|,\hspace{.5cm} \alpha_1=\pm1,\,\alpha_2=\pm1
\end{equation}
which is exactly the same as for a pseudospin-3/2. Note that a similar model has been studied in Ref. \cite{Roy_2018}.

The model (\ref{HAandB}) is described by a Hamiltonian vector
\begin{equation}
	\begin{aligned}
\mathbf{h}(\mathbf{q})&=\left(q_y+\frac{q_z}{2},0,0,0,q_x+\frac{q_y}{2},\frac{q_x}{2}+q_z,0,0,\right.\\
&\left.\hspace{.6cm}\frac{q_x}{2}-q_z,0,0,-q_x+\frac{q_y}{2},q_y-\frac{q_z}{2},0,0\right),
\end{aligned}
\end{equation}
from which the vectors $\mathbf{h}_\star$ and $\mathbf{h}_{\star\star}$ are readily obtained. The Bloch vectors follow from Eq. (\ref{balpha34}) as
\begin{equation}
	\begin{aligned}
\mathbf{b}_{\alpha_1\alpha_2}(\mathbf{q})&=\frac{\alpha_1}{1+2\alpha_2}\left[\alpha_2\frac{\mathbf{h}}{|\mathbf{q}|}\right.\\
&\left.\hspace{1.8cm}+\alpha_1\left(1+\frac{\alpha_2}{2}\right)\frac{\mathbf{h}_\star}{|\mathbf{q}|^2}+\frac{\mathbf{h}_{\star\star}}{|\mathbf{q}|^3}\right].
\end{aligned}
\end{equation}
The QGT for all four bands (calculated in the same way as for the spin-3/2 case above) is then given by
\begin{equation}
	\begin{aligned}
		\boldsymbol{\Omega}_{\alpha_1\alpha_2}(\mathbf{q})&=-\alpha_1(1+\alpha_2)\frac{\mathbf{q}}{2|\mathbf{q}|^3},\\
		g_{\alpha_1\alpha_2,ij}(\mathbf{q})&=\frac{1}{2|\mathbf{q}|^2}\left(\delta_{ij}-\frac{q_iq_j}{|\mathbf{q}|^2}\right).
	\end{aligned}
	\label{notspin}
\end{equation}
Though Eq. (\ref{notspin}) is formally similar to the QGT of a spin-3/2 fermion, there are crucial differences. Most importantly, the first Chern numbers obtained from Eq. (\ref{notspin}) are $\{2,0,0,-2\}$ from the lowest to the highest band, while they are $\{3,1,-1,-3\}$ for a spin-3/2. Moreover, the quantum metric is band-independent.

\section{Conclusions}
\label{conclude}

For a physical system described by a (Hermitian) parametric Hamiltonian matrix $H=H(\mathbf{x})$, there are situations in which it is useful to recall that any eigenstate is more fundamentally encoded in the \emph{eigenprojector matrix} $P_\alpha(\mathbf{x})=\ket{\psi_\alpha(\mathbf{x})}\bra{\psi_\alpha(\mathbf{x})}$ \footnote{We here use the term 'more fundamental' in the sense that, while $\ket{\psi_\alpha}$ and $P_\alpha$ can always be constructed from one another, $P_\alpha$ does not suffer from gauge arbitrariness and uncontrollable singularities.}. 

 Consequently, for a given quantity of interest whose standard expression is known in terms of eigenstates, it can prove rewarding to aim for a reformulation in terms of eigenprojectors, which will entirely eliminate the necessity for constructing eigenstates explicitly. 
Such a reformulation can be done for any physical quantity in principle. The present work demonstrates this by focusing on the case where the quantities of interest are the quantum metric and Berry curvature tensors (forming together what is known as the quantum geometric tensor, QGT). 
Selecting the QGT is natural for two reasons. First, it is relatively simple (compared to more involved quantities like the orbital magnetic susceptibility), and therefore serves well for illustrating the main features of the eigenprojector approach, which are preserved in the treatment of more complicated observables. Second, the inconvenient features of eigenstates become particularly problematic when one is interested in geometrical quantities; therefore, the QGT is one of the observables for which the projector formalism is the most powerful and beneficial. 

The most striking qualitative conclusion that can be drawn from the eigenprojector approach is that any quantity of interest can be written in terms of the Hamiltonian vector $\mathbf{h}$ and the eigenvalues $E_\alpha$, without using eigenstates, in agreement with recent results of Pozo and de Juan \cite{Pozo_2020} (who used a somewhat different mathematical framework). At the origin of this conclusion is the fact that each eigenprojector is a matrix polynomial of degree $N-1$ in $H$, according to the Cayley-Hamilton theorem.
 This result, though implicitly known for a long time \cite{Halmos_2017}, is perhaps most useful to physicists when formulated as an explicit function $P_\alpha(E_\alpha,H)$ of a single eigenvalue, see Eqs. (\ref{Projectormain}) \& (\ref{Project34}). These expressions for $P_\alpha(E_\alpha,H)$ represent our first intermediate result. 

The realization that the knowledge of $H$ as well as its eigenenergies immediately yields the eigenprojectors implies that any physical quantity of interest can be computed in the following way: 
\begin{enumerate}[label=(\arabic*),ref={(\arabic*)}]
\item Express the quantity in terms of eigenprojectors.
\item Explicitly insert the function $P_\alpha(E_\alpha,H)$. 
\end{enumerate}

In the course of carrying out this procedure, it proves extremely enlightening to switch between two "languages" where appropriate. The first language simply employs the relevant matrices ($H$, $P_\alpha$, \etc), while the second one expands those matrices in the generators of SU($N$). In this second (vectorial) language the Hamiltonian $H$ is given by a real vector $\mathbf{h}$,
 and similarly the eigenprojector $P_\alpha$ is represented by a real \emph{(generalized) Bloch vector} $\mathbf{b}_\alpha$. The analog of the matrix function $P_\alpha(E_\alpha,H)$ is the vector function $\mathbf{b}_\alpha(E_\alpha,\mathbf{h})$, see in particular the  important result (\ref{balpha34}). Both languages contain exactly the same information, such that the above protocol can be equivalently formulated as: 
\begin{enumerate}[label=(\arabic*),ref={(\arabic*)}]
	\item Express the quantity in terms of Bloch vectors.
	\item Explicitly insert the function $\mathbf{b}_\alpha(E_\alpha,\mathbf{h})$. 
\end{enumerate}

In the present paper, starting in Section \ref{blochgeo}, the above protocol was carried out for the QGT, and the systematic application to other, more complicated observables will be published elsewhere \cite{Graf}. The known formula (\ref{QGT23}) for the QGT in terms of eigenprojectors can be written in the vectorial language as described in Section \ref{altpics} [cf. in particular Eq. (\ref{geom})], which proves quite handy for practical computations and completes step (1). Step (2) was carried out in Section \ref{explsec}, where we obtained the Berry curvature tensor $\Omega_{\alpha,ij}$ in terms of the vector $\mathbf{h}$ and the relevant energy eigenvalue $E_\alpha$ only. This generalizes the well-known Berry curvature expression (\ref{Q2}) to arbitrary $N$.

As a concrete illustration of the results obtained for the QGT, Section \ref{examples} served to present models belonging to the class of multifold (Dirac) fermions, characterized by an $N$-fold degenerate nodal point accompanied by a linear band crossing.
First, we found interesting Bloch vector expressions for spin-1 and spin-$3/2$ fermions that yield a QGT consistent with the literature. Second, we introduced two multifold fermion models, which emphasize that different Hamiltonians can exhibit completely distinct geometrical and topological properties despite being indistinguishable according to their band structure. Such Hamiltonians may serve as interesting platforms for studying effects such as orbital magnetism or Friedel oscillations, similar to the $\alpha-T_3$ model \cite{Raoux_2014}. 

To close this paper, we state a few immediate perspectives of our work.
First, as already mentioned, it is worthwhile to apply the
eigenprojector approach to observables that are not purely of geometric
origin. In particular, the analytical insight afforded by this approach will help to unveil and distinguish purely
spectral and geometrical contributions to such observables, similar in
spirit to what has been done for the orbital magnetic susceptibility of
two-band models in Ref. \cite{Raoux_2015}, superconducting stiffness or non-linear responses and more generalized thermodynamic stiffnesses.

Second, it would be interesting to establish a better understanding of the geometry and topology of multifold fermion models with linear but also quadratic band crossings. More concretely, how to systematically design models that all share a certain type of energy spectrum but belong to different topological classes according to the
tenfold way classification \cite{Ludwig_2015}?

Finally, note that the validity of the formalism developed in this paper
goes beyond the case of Hermitian Hamiltonian matrices. In particular, the key expressions Eq. (\ref{Project34}) for the eigenprojector
$P_\alpha(E_\alpha,H)$ and Eq. (\ref{balpha34}) for the Bloch vector
$\mathbf{b}_\alpha(E_\alpha,\mathbf{h})$ stay valid for systems where the Hermiticity condition
is relaxed \cite{Bender_2007,Brody_2013}. 

\section*{Acknowledgements}
	We thank Mark-Oliver Goerbig and Andrej Mesaros for fruitful discussions.

\appendix

\begin{widetext}
	
\section{Gauge dependency and singularities of eigenstates}
	\label{App2states}
	
Here, we describe some issues of eigenstates that are absent in an eigenprojector approach. Consider the two-band eigenstates (\ref{2states}).	The eigenstate for $\alpha=+$ can be written in either of the forms
	\begin{equation}
		\begin{aligned}
			\ket{\psi_+(\mathbf{x})}&=e^{i\Gamma_+}\begin{pmatrix}
				\cos\theta_+\\
				\sin\theta_+ e^{i\phi_+}
			\end{pmatrix}=
			\frac{1}{\sqrt{2\left(1+\frac{h_z}{E_+}\right)}}\begin{pmatrix}
				1+\frac{h_z}{E_+}\\
				\frac{h_x+ih_y}{E_+}
			\end{pmatrix}=\frac{1}{\sqrt{2\left(1+\cos\theta_h\right)}}\begin{pmatrix}
				1+\cos\theta_h\\
				\sin\theta_h e^{i\phi_h}
			\end{pmatrix},
		\end{aligned}
		\label{2psi}
	\end{equation}
	and similarly for $\ket{\psi_-(\mathbf{x})}$.
	The first expression corresponds to the formal parametrization of a complex-valued two-component unit vector. It expresses the fact that, at each point $\mathbf{x}$ in the parameter space, the eigenstate $\ket{\psi_+(\mathbf{x})}$ is minimally encoded by a global phase $\Gamma_+(\mathbf{x})$
	and two eigenstate's angles $(\theta_+(\mathbf{x}),\phi_+(\mathbf{x}))$. Thus, the first evident problem of Eq. (\ref{2psi}) consists in the fact that it is gauge-dependent.
	The second expression in Eq. (\ref{2psi}) illustrates a second problem, namely the possible singular behavior of eigenstates. A singularity is evidently to be expected when $h_z(\mathbf{x}_0)/E_+(\mathbf{x}_0)=-1$ for some point $\mathbf{x}=\mathbf{x}_0$ in parameter space. Similarly, if the eigenstate components are written in terms of the angles $(\theta_h,\phi_h)$, the explicit relation $\tan \phi_h=h_y/h_x$ points towards further possible singular behavior 
	in parameter space when $h_x(\mathbf{x}_0)=0$. Finally, the closed-form expression of $\ket{\psi_+(\mathbf{x})}$ (in terms of the corresponding energy eigenvalue and the components of the Hamiltonian vector $\mathbf{h}$) is rather cumbersome and not very convenient if one is dealing with classes of Hamiltonians beyond a specific Hamiltonian of interest. This issue in particular becomes much more dramatic as soon as $N>2$. In this case, similar closed-form expressions are practically useless.

	\section{Quantum geometric tensor for different parameter spaces}
	\label{Appparamchoice}
	
	The explicit dimension and form of the matrices $g_\alpha$ and $\Omega_\alpha$ clearly depends on the parameter space. For example, in a two-band system, taking $\mathbf{x}=(\theta_h,\phi_h)$ ($d=\dim(\mathbf{x})=2$) and 
	inserting the state (\ref{2psi}) into Eq. (\ref{QG0}), 
	the quantum metric and Berry curvature tensors are $2\times 2$ matrices
	that write as
	\begin{equation}
		\begin{aligned}
			g_{+}(\theta_h,\phi_h)=\frac{1}{4}\begin{pmatrix}
				1 & 0\\
				0 & \sin^2\theta_h
			\end{pmatrix}, && \Omega_{+}(\theta_h,\phi_h)=-\frac{1}{2}
			\begin{pmatrix}
				0 & \sin\theta_h\\
				-\sin\theta_h &0
				\label{2T}
			\end{pmatrix},
		\end{aligned}
	\end{equation}
	which is a well-known result \cite{Kolodrubetz_2017}. 
	Taking instead $\mathbf{x}=(h_x,h_y,h_z)$ ($d=\dim(\mathbf{x})=3$) for the same two-band state, the quantum metric and Berry curvature tensors are now $3\times3$ matrices with matrix elements
	\begin{equation}
		\begin{aligned}
			g_{+,ij}(\mathbf{h})=\frac{1}{4|\mathbf{h}|^2}\left(\delta_{ij}-\frac{h_i h_j}{|\mathbf{h}|^2}\right),&&
			\Omega_{+,ij}(\mathbf{h})=-\frac{1}{2|\mathbf{h}|^3}\epsilon_{ijk}h_k,
			\label{2Tb}
		\end{aligned}
	\end{equation}
	where $\epsilon_{ijk}$ is the Levi-Civita antisymmetric tensor. Note that $g_{-,ij}=g_{+,ij}$ and $\Omega_{-,ij}=-\Omega_{+,ij}$ for the state $\ket{\psi_-}$.
	
	More generally, the QGT of interest is often related to the explicit dependency of the Hamiltonian vector $\mathbf{h}(\mathbf{x})$
	on some vector of external parameters $\mathbf{x}$, with $d=\dim(\mathbf{x})\ge 2$. For example, in condensed matter physics, one will often have $\mathbf{x}=\mathbf{k}$, where $\mathbf{k}$ represents crystal momentum. In that situation,
	the corresponding QGT $T_{\alpha,ij}(\mathbf{x})$ may be obtained from either $T_{\alpha,kl}(\mathbf{h})$ or $T_{\alpha,kl}(\theta_h,\phi_h)$ by a simple composition rule as
	\begin{equation}
		T_{\alpha,ij}(\mathbf{x})=\sum_{k,l}(\partial_i y_k)(\partial_j y_l)T_{\alpha,kl}(\mathbf{y}),
		\label{trans}
	\end{equation}
	with $\mathbf{y}=\mathbf{h}$ or $\mathbf{y}= (\theta_h,\phi_h)$.
	
	\section{Gell-Mann and spin matrices}
	\label{GellMann}
	
	Here we list the $N=3$ ($N=4$) Gell-Mann matrices and relate them to spin-1 (spin-3/2) matrices. Note that the generalization to $N\geq5$ Gell-Mann matrices (not listed here explicitly) is straightforward, see for example Refs. \cite{Pfeifer_2003,Bertlmann_2008}: There are $N(N-1)/2$ symmetric matrices (purely real), $N(N-1)/2$ antisymmetric matrices (purely imaginary), and $N-1$ diagonal matrices.
	
	The $N=3$ Gell-Mann matrices \cite{GellMann_1962} are given by
	\[
	\begin{aligned}
		\lambda_1&=
		\begin{pmatrix}
			0 & 1 & 0\\ 1 & 0 & 0 \\ 0 & 0 & 0
		\end{pmatrix},
		&
		\lambda_2&=
		\begin{pmatrix}
			0 & -i & 0\\ i & 0 & 0 \\ 0 & 0 & 0
		\end{pmatrix},
		&
		\lambda_3&=
		\begin{pmatrix}
			1 & 0 & 0\\ 0 & -1 & 0 \\ 0 & 0 & 0
		\end{pmatrix},
		&
		\lambda_4&=
		\begin{pmatrix}
			0 & 0 & 1\\ 0 & 0 & 0 \\ 1 & 0 & 0
		\end{pmatrix},\\
		\lambda_5&=
		\begin{pmatrix}
			0 & 0 & -i\\ 0 & 0 & 0 \\ i & 0 & 0
		\end{pmatrix},
		&
		\lambda_6&=
		\begin{pmatrix}
			0 & 0 & 0\\ 0 & 0 & 1 \\ 0 & 1 & 0
		\end{pmatrix},
		&
		\lambda_7&=
		\begin{pmatrix}
			0 & 0 & 0\\ 0 & 0 & -i \\ 0 & i & 0
		\end{pmatrix},
		&
		\lambda_8&=
		\frac{1}{\sqrt{3}}\begin{pmatrix}
			1 & 0 & 0\\ 0 & 1 & 0 \\ 0 & 0 & -2
		\end{pmatrix}.
	\end{aligned}
	\]		
	They can easily be related to spin-1 matrices, which in the standard representation read
	\begin{equation}
		\begin{aligned} S_x=\frac{1}{\sqrt{2}}(\lambda_1+\lambda_6),&& S_y=\frac{1}{\sqrt{2}}(\lambda_2+\lambda_7),&& S_z=\frac{1}{2}(\lambda_3+\sqrt{3}\lambda_8).
		\end{aligned}
		\label{standardrep}
	\end{equation}
	
	The $N=4$ Gell-Mann matrices \cite{Pfeifer_2003,Bertlmann_2008} are given by the extended SU($3$) Gell-Mann matrices,
	\[
	\begin{aligned}
		\lambda_1&=
		\begin{pmatrix}
			0 & 1 & 0 & 0 \\ 1 & 0 & 0 & 0 \\ 0 & 0 & 0 & 0 \\ 0 & 0 & 0 & 0
		\end{pmatrix},
		&
		\lambda_2&=
		\begin{pmatrix}
			0 & -i & 0 & 0 \\ i & 0 & 0 & 0 \\ 0 & 0 & 0 & 0 \\ 0 & 0 & 0 & 0
		\end{pmatrix},
		&
		\lambda_3&=
		\begin{pmatrix}
			1 & 0 & 0 & 0 \\ 0 & -1 & 0 & 0 \\ 0 & 0 & 0 & 0 \\ 0 & 0 & 0 & 0
		\end{pmatrix},&
		\lambda_4&=
		\begin{pmatrix}
			0 & 0 & 1 & 0 \\ 0 & 0 & 0 & 0 \\ 1 & 0 & 0 & 0 \\ 0 & 0 & 0 & 0
		\end{pmatrix},\\
		\lambda_5&=
		\begin{pmatrix}
			0 & 0 & -i & 0 \\ 0 & 0 & 0 & 0 \\ i & 0 & 0 & 0 \\ 0 & 0 & 0 & 0
		\end{pmatrix},
		&
		\lambda_6&=
		\begin{pmatrix}
			0 & 0 & 0 & 0 \\ 0 & 0 & 1 & 0 \\ 0 & 1 & 0 & 0 \\ 0 & 0 & 0 & 0
		\end{pmatrix},&
		\lambda_7&=
		\begin{pmatrix}
			0 & 0 & 0 & 0 \\ 0 & 0 & -i & 0 \\ 0 & i & 0 & 0 \\ 0 & 0 & 0 & 0
		\end{pmatrix},
		&
		\lambda_8&=\frac{1}{\sqrt{3}}\begin{pmatrix}
			1 & 0 & 0 & 0 \\ 0 & 1 & 0 & 0 \\ 0 & 0 & -2 & 0 \\ 0 & 0 & 0 & 0
		\end{pmatrix},\\
	\end{aligned}\]
	plus an additional seven matrices
	\[
	\begin{aligned}
		\lambda_9&=
		\begin{pmatrix}
			0 & 0 & 0 & 1 \\ 0 & 0 & 0 & 0 \\ 0 & 0 & 0 & 0 \\ 1 & 0 & 0 & 0
		\end{pmatrix},
		&
		\lambda_{10}&=
		\begin{pmatrix}
			0 & 0 & 0 & -i \\ 0 & 0 & 0 & 0 \\ 0 & 0 & 0 & 0 \\ i & 0 & 0 & 0
		\end{pmatrix},
		&
		\lambda_{11}&=
		\begin{pmatrix}
			0 & 0 & 0 & 0 \\ 0 & 0 & 0 & 1 \\ 0 & 0 & 0 & 0 \\ 0 & 1 & 0 & 0
		\end{pmatrix},&
		\lambda_{12}&=
		\begin{pmatrix}
			0 & 0 & 0 & 0 \\ 0 & 0 & 0 & -i \\ 0 & 0 & 0 & 0 \\ 0 & i & 0 & 0
		\end{pmatrix},\\
		\lambda_{13}&=
		\begin{pmatrix}
			0 & 0 & 0 & 0 \\ 0 & 0 & 0 & 0 \\ 0 & 0 & 0 & 1 \\ 0 & 0 & 1 & 0
		\end{pmatrix},
		&
		\lambda_{14}&=\begin{pmatrix}
			0 & 0 & 0 & 0 \\ 0 & 0 & 0 & 0 \\ 0 & 0 & 0 & -i \\ 0 & 0 & i & 0
		\end{pmatrix},&
		\lambda_{15}&=\frac{1}{\sqrt{6}}\begin{pmatrix}
			1 & 0 & 0 & 0 \\ 0 & 1 & 0 & 0 \\ 0 & 0 & 1 & 0 \\ 0 & 0 & 0 & -3
		\end{pmatrix}.
	\end{aligned}
	\]
	This can easily be used to express the standard representation of spin-3/2 matrices: 
	\begin{equation}
		\begin{aligned}
			S_x&=\frac{1}{2}(\sqrt{3}\lambda_1+2\lambda_6+\sqrt{3}\lambda_{13}),\\ S_y&=\frac{1}{2}(\sqrt{3}\lambda_2+2\lambda_7+\sqrt{3}\lambda_{14}),\\ S_z&=\frac{1}{2}(\lambda_3+\sqrt{3}\lambda_8+\sqrt{6}\lambda_{15}).
			\label{spin4}
		\end{aligned}
	\end{equation}

\section{Derivation of Eq. (\ref{Projectormain})}
\label{AppEle}

To find the eigenprojector $P_\alpha$ as a polynomial in the Hamiltonian, our starting point is the following textbook formula, involving the set $\{E_\beta,\beta=1,...,N\}$ of all energy eigenvalues \cite{Halmos_2017}:
\begin{equation}
	P_\alpha=\prod_{\beta\neq\alpha}\frac{H-E_\beta 1_N}{E_\alpha-E_\beta}.
	\label{projknown}
\end{equation}
Note that in the language of matrix theory, Eq. (\ref{projknown}) defines the \emph{Frobenius covariants} of $H$. Note also that the denominator of Eq. (\ref{projknown}) corresponds to the derivative of the Hamiltonian's characteristic polynomial. More generally, before proceeding it proves useful to compile some more details on the characteristic polynomial $p_N(z)$.

\subsection{Characteristic polynomial}
\label{enswaves}

The \emph{characteristic polynomial} of an $N\times N$ matrix $A$ is given by \cite{Gantmacher_1980}
\begin{equation}
	\tilde{p}_N(z)=\det(z 1_N-A)=\sum_{k=0}^N \tilde{c}_k z^{N-k}=\prod_{\alpha=1}^N(z-a_\alpha),
\end{equation}
where $a_\alpha$ denotes an eigenvalue of $A$ and $\tilde{c}_0=1$. According to the Faddeev-Le Verrier algorithm \cite{Gantmacher_1980}, the coefficients $\tilde{c}_k$ may be computed from the traces $s_k\equiv\Tr A^k$ of powers of $A$ as
\begin{equation}
	\begin{aligned}
		\tilde{c}_k&=-\frac{1}{k}(s_k+\tilde{c}_1s_{k-1}+...+\tilde{c}_{k-1}s_1)\\
		&=\frac{(-1)^k}{k!}Y_k(s_1,...,(-1)^{k-1}(k-1)!s_k).
	\end{aligned}
\end{equation}
The second equality involves \emph{(exponential) complete Bell polynomials} $Y_k(z_1,...,z_k)$ \cite{Bell_1934,Comtet_1974}, the first few of which read explicitly
\begin{equation}
	\begin{aligned}
		Y_0&=1,\\
		Y_1(z_1)&=z_1,\\
		Y_2(z_1,z_2)&=z_1^2+z_2,\\
		Y_3(z_1,z_2,z_3)&=z_1^3+3z_1z_2+z_3,\\
		Y_4(z_1,z_2,z_3,z_4)&=z_1^4+6z_1^2z_2+4z_1z_3+3z_2^2+z_4.
	\end{aligned}
\end{equation}
Focusing now on the case where $A=H$ represents an $N\times N$ Hamiltonian matrix, we have the Hamiltonian's characteristic polynomial
\begin{equation}
	p_N(z)=\det(z 1_N-H)=\sum_{k=0}^N c_k z^{N-k}=\prod_{\alpha=1}^N(z-E_\alpha).
	\label{charpol}
\end{equation}
Using the traceless character of the Hamiltonian, cf. Eq. (\ref{genham0}), and the Casimir invariants defined in Eq. (\ref{Cndef}), the coefficients are given by 
\begin{equation}
	c_k=\frac{(-1)^k}{k!}Y_k(0,-C_2,...,(-1)^{k-1}(k-1)!C_k),
	\label{charcoeffs}
\end{equation}
and the first few of them read explicitly
\begin{equation}
	\begin{aligned}
		c_0&=1,&& &c_1&=0,\\
		c_2&=-\frac{C_2}{2},&& &c_3&=-\frac{C_3}{3},\\
		c_4&=\frac{C_2^2}{8}-\frac{C_4}{4},&& &c_5&=\frac{C_2C_3}{6}-\frac{C_5}{5},\\
		c_6&=-\frac{C_2^3}{48}+\frac{C_3^2}{18}+\frac{C_2C_4}{8}-\frac{C_6}{6}. &&
		\label{traces}
	\end{aligned}
\end{equation}

\subsection{Rewriting Eq. (\ref{projknown})}

The goal is now to eliminate all $E_{\beta\neq\alpha}$ from Eq. (\ref{projknown}), 
such that $P_\alpha$ becomes a proper polynomial in $H$ with coefficients that depend only on the single eigenvalue $E_\alpha$, \ie $P_\alpha=P_\alpha(E_\alpha,H)$.
Consider first the numerator of Eq. (\ref{projknown}) and note that by explicit multiplication one may write
\begin{equation}
	\prod_{\beta\neq\alpha}(H-E_\beta1_N)=\sum_{n=0}^{N-1}(-1)^ne_n(E_1,...,E_{\alpha-1},E_{\alpha+1},...,E_N)H^{N-1-n},
	\label{step1}
\end{equation}
where $e_n=e_n(E_1,...,E_{\alpha-1},E_{\alpha+1},...,E_N)$ are known as \emph{elementary symmetric polynomials} \cite{Macdonald_1998}. One has $e_0=1$ and all higher $e_n$ are determined recursively by Newton's identities:
\begin{equation}
	e_n=\frac{1}{n}\sum_{k=1}^{n}(-1)^{k-1}(C_k-E_\alpha^k)e_{n-k},
\end{equation}
where the $C_k$ are the Casimir invariants of Eq. (\ref{Cndef}) and it was exploited that $\sum_{\beta\neq\alpha}E_\beta^k=C_k-E_\alpha^k$.
This may further be rewritten as
\begin{equation}
	e_n=(-1)^n\sum_{k=0}^nc_kE_\alpha^{n-k},
\end{equation}
where $c_k$ are the coefficients (\ref{charcoeffs}) of the characteristic polynomial.
If we now define polynomials 
\begin{equation}
	q_n(z)\equiv\sum_{k=0}^{n}c_kz^{n-k},
	\label{qs}
\end{equation}
it is clear that $q_N(z)=p_N(z)$ is the characteristic polynomial (\ref{charpol}), and $q_n(E_\alpha)=(-1)^ne_n$. Moreover,
inserting into Eq. (\ref{step1}), we have
\begin{equation}
	\prod_{\beta\neq\alpha}(H-E_\beta1_N)=\sum_{n=0}^{N-1}q_{N-1-n}(E_\alpha)H^n.
\end{equation}
Similarly, for the numerator of Eq. (\ref{projknown}), exactly the same procedure as above (where $H$ gets replaced by $E_\alpha$) leads to 
\begin{equation}
	\prod_{\beta\neq\alpha}(E_\alpha-E_\beta)=\sum_{n=0}^{N-1}q_{N-1-n}(E_\alpha)E_\alpha^n.
\end{equation}
As mentioned in the main text, $\prod_{\beta\neq\alpha}(E_\alpha-E_\beta)$ is equal to the derivative $p_N'(E_\alpha)$ of the characteristic polynomial. From this one may also show that $\prod_{\beta\neq\alpha}(E_\alpha-E_\beta)=\sum_{n=0}^{N-1}q_{N-1-n}(E_\alpha)C_n$. Combining all of these results, one arrives at Eq. (\ref{Projectormain}).

	\section{Eigenstates from eigenprojectors}
	\label{ApppsifromP}
	
The eigenprojector $P_\alpha$ permits to construct the eigenstate $\ket {\psi_\alpha}$ from an arbitrary gauge freedom state $\ket {\psi_g}$. 
	For example, in the two-band case, using the eigenprojector $P_+$ given by Eq. (\ref{Project34}), a state equivalent to Eq. (\ref{2psi}) is easily constructed as
	\begin{equation}
		\begin{aligned}
			\ket {\psi_+}=\frac{1}{\sqrt{\bra{\psi_g} P_+  \ket {\psi_g}}} P_+  \ket {\psi_g}, &&
			\ket {\psi_g}=\begin{pmatrix}
				\cos \theta_g \\
				\sin \theta_g  e^{-i\phi_g} 
			\end{pmatrix},
			\label{psi2}
		\end{aligned}
	\end{equation}
	where the gauge freedom angles $(\theta_g,\phi_g)$ can be chosen at will at any point in parameter space $\mathbf{x}$. 
	For example, if $\cos\theta_g=1$ and $\sin\theta_g=0$, one exactly recovers the second expression in  Eq. (\ref{2psi}).
	More generally, the $N$-band eigenstate $\ket {\psi_\alpha}$ in an arbitrary gauge may be obtained as
	\begin{equation}
		\begin{aligned}
			\ket {\psi_\alpha}=\frac{1}{\sqrt{\bra{\psi_g} P_\alpha(E_\alpha,H)  \ket {\psi_g}}} P_\alpha(E_\alpha,H)  \ket {\psi_g},
		\end{aligned}
		\label{psiN}
	\end{equation}
	where the gauge freedom state $\ket {\psi_g}$ can be chosen arbitrarily. 
	Each eigenstate (\ref{psiN}) is minimally encoded by a global phase $\Gamma_\alpha$ and 
	$N-1$ pairs of eigenstate's angles $(\theta_\alpha ^i,\phi_\alpha ^i)$ ($i=1,...,N-1$). 
	
Two more remarks can be made.
	First, although this is not immediately obvious, the eigenstate's angles
	stay unchanged upon varying $\ket{\psi_g}$ for fixed $\mathbf{x}$. In contrast, the global phase is changing, 
	implying $\Gamma_\alpha\equiv \Gamma_\alpha( \mathbf{x},\ket{\psi_g})$. As a consequence, the possible singular behaviors of the wavefunction in parameter space are gauge-dependent.
	Second, the state $\ket {\psi_g}$ must not be orthogonal to the projector $P_\alpha(\mathbf{x})$, \ie one requires $P_\alpha(\mathbf{x})\ket {\psi_g}\ne 0$. This constraint implies	
	that it might be necessary to change the state $\ket {\psi_g}$ when the parameter $\mathbf{x}$ is varying because 
	it is never guaranteed that a single $\ket {\psi_g}$ (meaning a fixed gauge) is sufficient to describe a given eigenstate $\ket {\psi_\alpha}$
	over the entire parameter space $\mathbf{x}$.

\section{Derivation of Bloch vector formula $\mathbf{b}_\alpha(E_\alpha,H)$}
\label{Appblochder}

Here we detail the steps that allow to find $\mathbf{b}_\alpha(\mathbf{h}, E_\alpha)$
from the eigenprojector formula (\ref{Projectormain}). The only missing ingredient 
is $H^n=(\mathbf{h} \cdot \boldsymbol{\lambda})^n$ in terms of the generators of SU($N$). Defining vectors $\boldsymbol{\eta}_n\equiv\Tr(H^n\boldsymbol{\lambda})/2$, we can write:
\begin{equation}
	H^n=(\mathbf{h}\cdot\boldsymbol{\lambda})^n=\frac{C_n}{N}1_N+\boldsymbol{\eta}_n\cdot\boldsymbol{\lambda},
	\label{powers}
\end{equation}
where obviously $\boldsymbol{\eta}_0=0$ and $\boldsymbol{\eta}_1=\mathbf{h}$.
Inserting into Eq. (\ref{Projectormain}) and using Eq. (\ref{defproj}), we obtain the intermediate result:
\begin{equation}
	\begin{aligned}
		\mathbf{b}_\alpha
		&=2\frac{\sum_{n=0}^{N-1}q_{N-1-n}(E_\alpha)\boldsymbol{\eta}_n}{\sum_{n=0}^{N-1}q_{N-1-n}(E_\alpha)E_\alpha^n}.
	\end{aligned}
	\label{Projectormain2}
\end{equation}

At this point it remains the task to find the explicit form of the vectors $\boldsymbol{\eta}_n(\mathbf{h})$ for $n>1$. 
To accomplish this task, we require a product identity generalizing the familiar SU($2$) identity (\ref{pauliprod}) to any $N$. This can be readily achieved by using Eq. (\ref{comms}) to obtain
\begin{equation}
	\lambda_a\lambda_b=\frac{2}{N}\delta_{ab}1_N+(d_{abc}+if_{abc})\lambda_c.
	\label{algebra}
\end{equation}
Since all generators are traceless, $\Tr\lambda_c=0$, the convenient trace orthogonality $\Tr(\lambda_a\lambda_b)=2\delta_{ab}$ holds. From Eqs. (\ref{algebra}) \& (\ref{prods}), 
one directly obtains the desired product identity in vector form:
\begin{equation}
	(\mathbf{m}\cdot\boldsymbol{\lambda})(\mathbf{n}\cdot\boldsymbol{\lambda})=\frac{2}{N}\mathbf{m}\cdot\mathbf{n}\,
	1_N+(\mathbf{m}\star\mathbf{n}+i\,\mathbf{m}\times\mathbf{n})\cdot\boldsymbol{\lambda}.
	\label{product}
\end{equation}
When computing $H^n=(\mathbf{h}\cdot\boldsymbol{\lambda})^n$ by applying Eq. (\ref{product}) repeatedly, it proves useful to introduce the notation (\ref{stardef}) for repeated star products of a vector with itself.
The resulting vectors have the following properties (with $n_1,n_2\in\mathbb{N}_0$):
\begin{equation}
	\begin{aligned}
		\mathbf{m}_\star^{(n_1)}\cdot\mathbf{m}_\star^{(n_2)}&=\left|\mathbf{m}_\star^{\left(\frac{n_1+n_2}{2}\right)}\right|^2
		& \text{ if }n_1+n_2 \text{ even},\\
		\mathbf{m}_\star^{(n_1)}\times\mathbf{m}_\star^{(n_2)}&=0. &
	\end{aligned}
	\label{ortho}
\end{equation}
The former identity follows directly from the total symmetry of the structure constants (\ref{comms}), 
and the latter is a consequence of the second Jacobi identity, as can be proved by mathematical induction. More generally, from the generic properties of the algebra, we have established useful generalized {\em vector} and {\em scalar} Jacobi identities, listed in Appendix \ref{AppJacob}.

With all these prerequisites we may now calculate ($\mathbf{h}\cdot\boldsymbol{\lambda})^n$\ \ie determine $C_n(\mathbf{h})$ and $\boldsymbol{\eta}(\mathbf{h})$. In particular, we obtain the following simple recursion relations:
\begin{equation}
	\begin{aligned}
		C_{n+1}=2\, \mathbf{h}\cdot\boldsymbol{\eta}_n,& &&&&
		\boldsymbol{\eta}_{n+1}=\mathbf{h}\star \boldsymbol{\eta}_{n}+\frac{C_n}{N} \mathbf{h},
	\end{aligned}
	\label{CIrecurs}
\end{equation}
with initial conditions $C_0=N$ and $\boldsymbol{\eta}_0=0$. Making use of the fact that $\boldsymbol{\eta}_n$ and the structure constants $f_{abc}$ are real by definition, we can also establish the identity
\begin{equation}
	\boldsymbol{\eta}_{n_1}\times\boldsymbol{\eta}_{n_2}=0,\,\,\,\,\forall\, n_1,n_2\in\mathbb{N}_0.
	\label{orth}
\end{equation}
 Applying the recursion (\ref{CIrecurs}), for a traceless Hamiltonian matrix, we obtain successively up to $n=4$ the important identities
\begin{equation}
	\begin{aligned}
		C_1&=0, & \boldsymbol{\eta}_1&=\mathbf{h},\\
		C_2&=2|\mathbf{h}|^2, & \boldsymbol{\eta}_2&=\mathbf{h}_\star, \\
		C_3&=2\,\mathbf{h}\cdot\mathbf{h}_\star, & \boldsymbol{\eta}_3&=\frac{C_2}{N}\mathbf{h}+\mathbf{h}_{\star\star},\\
		C_4&=\frac{4|\mathbf{h}|^4}{N}+2|\mathbf{h}_\star|^2, & \boldsymbol{\eta}_4&=\frac{C_3}{N}\mathbf{h}+\frac{C_2}{N}\mathbf{h}_\star+\mathbf{h}_{\star\star\star}.\\
	\end{aligned}
	\label{invariants}
\end{equation}
More generally, for a generic $N>1$, the form of the vector $\boldsymbol{\eta}_n(\mathbf{h})$ is compactly written as:
\begin{equation}
	\begin{aligned}
		\boldsymbol{\eta}_{n}=\frac{1}{N}\sum_{p=0} ^{n-1} C_p \mathbf{h}_{\star} ^{(n-1-p)}.
	\end{aligned}
	\label{etan}
\end{equation}
The final step required for completing our task of finding the explicit expressions of the Bloch vector $\mathbf{b}_\alpha(\mathbf{h},E_\alpha)$ 
consists in substituting Eqs. (\ref{invariants}) \& (\ref{etan}) into Eq. (\ref{Projectormain2}). One can then write down the explicit Bloch vectors for any $N$, as done explicitly in Eq. (\ref{balpha34}) for $N=2$ to $N=5$.

As a final remark, be aware that, for given $N$, only $C_{n\leq N}$ and $\boldsymbol{\eta}_{n<N}$ are relevant, and there are only $N-1$ 
independent vectors \smash{$\mathbf{h}_\star^{(k)}$}, $k=0,...,N-2$. For example, for $N=3$, all information we need is encoded in $C_2$, $C_3$, $\mathbf{h}$ and $\mathbf{h}_\star$. 
This again follows from the Cayley-Hamilton theorem, which states that $q_{n=N}(H)=p_N(H)=0$. From this property it is easy to establish the following useful identities:
\begin{equation}
	\begin{aligned}
		N&=3:\hspace{1cm} \mathbf{h}_{\star\star}=\frac{C_2}{6}\mathbf{h},\hspace{1cm} \mathbf{h}_\star\star\mathbf{h}_\star=\frac{C_3}{3}\mathbf{h}-\frac{C_2}{6}\mathbf{h}_\star,\\
		N&=4:\hspace{1cm} \mathbf{h}_{\star\star\star}=\frac{C_3}{12}\mathbf{h}+\frac{C_2}{4}\mathbf{h}_\star,\hspace{1cm} \mathbf{h}_\star\star\mathbf{h}_\star=\frac{C_3}{3}\mathbf{h},\hspace{1cm} \mathbf{h}_\star\star\mathbf{h}_{\star\star}=\frac{|\mathbf{h}_\star|^2}{2}\mathbf{h}+\frac{C_3}{12}\mathbf{h}_\star.
		\label{starrelsb}
	\end{aligned}
\end{equation}

\section{SU($N$) Jacobi identities}
\label{AppJacob}

The {\em first Jacobi identity}  \cite{Kaplan_1967, Macfarlane1968} can be written alternatively for the generator matrices, the antisymmetric structure constants and the SU($N$) vectors as
\begin{equation}
	\begin{aligned}
		\comm{\comm{\lambda_i}{\lambda_j}}{\lambda_k}+\comm{\comm{\lambda_j}{\lambda_k}}{\lambda_i}+\comm{\comm{\lambda_k}{\lambda_i}}{\lambda_j}&=0,\\
		f_{ijm}f_{klm}+f_{ikm}f_{ljm}+f_{ilm}f_{jkm}&=0,\\
		\mathbf{m}\times(\mathbf{n}\times\mathbf{o})+\mathbf{n}\times(\mathbf{o}\times\mathbf{m})+\mathbf{o}\times(\mathbf{m}\times\mathbf{n})&=0,\\
		(\mathbf{m}\times\mathbf{n})\cdot(\mathbf{o}\times\mathbf{p })+(\mathbf{m}\times\mathbf{o})\cdot(\mathbf{p}\times\mathbf{n})
		+(\mathbf{m}\times\mathbf{p})\cdot(\mathbf{n}\times\mathbf{o})&=0,
	\end{aligned}
\end{equation}
where the third and fourth lines are obtained from the second line depending on whether or not one keeps a free index.
Similarly, the {\em second Jacobi identity} is given by
\begin{equation}
	\begin{aligned}
		\comm{\acomm{\lambda_i}{\lambda_j}}{\lambda_k}+\comm{\acomm{\lambda_j}{\lambda_k}}{\lambda_i}+\comm{\acomm{\lambda_k}{\lambda_i}}{\lambda_j}&=0,\\
		f_{ijm}d_{klm}+f_{ikm}d_{ljm}+f_{ilm}d_{jkm}&=0,\\
		\mathbf{m}\times(\mathbf{n}\star\mathbf{o})+\mathbf{n}\times(\mathbf{o}\star\mathbf{m})+\mathbf{o}\times(\mathbf{m}\star\mathbf{n})&=0,\\
		(\mathbf{m}\times\mathbf{n})\cdot(\mathbf{o}\star\mathbf{p })+(\mathbf{m}\times\mathbf{o})\cdot(\mathbf{p}\star\mathbf{n})
		+(\mathbf{m}\times\mathbf{p})\cdot(\mathbf{n}\star\mathbf{o})&=0.
	\end{aligned}
	\label{jacobi2}
\end{equation}
Furthermore, there is the identity
\begin{equation}
	\begin{aligned}
		\comm{\lambda_i}{\comm{\lambda_j}{\lambda_k}}&=\acomm{\lambda_k}{\acomm{\lambda_i}{\lambda_j}}-\acomm{\lambda_j}{\acomm{\lambda_i}{\lambda_k}},\\
		f_{ijm}f_{klm}&=\frac{2}{N}(\delta_{ik}\delta_{jl}-\delta_{il}\delta_{jk})+d_{ikm}d_{jlm}-d_{ilm}d_{jkm},\\
		\mathbf{m}\times(\mathbf{n}\times\mathbf{o})&=
		\frac{2}{N}[(\mathbf{m}\cdot\mathbf{o})\mathbf{n}-(\mathbf{m}\cdot\mathbf{n})\mathbf{o}]
		+(\mathbf{m}\star\mathbf{o})\star\mathbf{n}-(\mathbf{m}\star\mathbf{n})\star\mathbf{o},\\
		(\mathbf{m}\times\mathbf{n})\cdot(\mathbf{o}\times\mathbf{p})&=
		\frac{2}{N}[(\mathbf{m}\cdot\mathbf{o})(\mathbf{n}\cdot\mathbf{p})-(\mathbf{m}\cdot\mathbf{p})(\mathbf{n}\cdot\mathbf{o})]+(\mathbf{m}\star\mathbf{o})\cdot(\mathbf{n}\star\mathbf{p})-(\mathbf{m}\star\mathbf{p})\cdot(\mathbf{n}\star\mathbf{o}).
		\label{jacobi3}
	\end{aligned}
\end{equation}

\section{The generalized Bloch sphere}
\label{geosphere}

We give here a short summary of the concept of a {\em generalized Bloch sphere} [\ie the SU($N$) eigenprojector's Bloch sphere \smash{$\mathcal{B}_{P_\alpha}^{(N)}$} introduced in Section \ref{genbs}], drawing largely on Refs. \cite{Harriman_1978,Jakobczyk_2001,Kimura_2003,Zyczkowski_2003,Byrd_2003,Kimura_2005,Mendas_2006,Goyal_2016}. 

For an $N$-dimensional Hilbert space, the three defining properties of a (mixed state) density matrix $\rho_\alpha$ representing an $N$-component (mixed) quantum state $\ket{\psi_\alpha}$, where $\alpha=1,...,N$, are hermiticity \smash{$\rho_\alpha^\dagger=\rho_\alpha$}, probability conservation $\Tr\rho_\alpha=1$ and positive semidefiniteness, $\rho_\alpha\geq0$. Pure states have, in addition, $\rho_\alpha^2=\rho_\alpha$, in which case $\rho_\alpha=P_\alpha$ is an eigenprojector. An expansion in the basis of SU($N$) generator matrices analogous to Eq. (\ref{defproj}) can be made for a general (mixed state) density matrix:
\begin{equation} 
\rho_\alpha=\frac{1}{N}1_N+\frac{1}{2}\mathbf{b}_\alpha\cdot\boldsymbol{\lambda},
\label{rhoeq}
\end{equation}
where now $|\mathbf{b}_\alpha|$ can take various values depending on the pureness of the state. In the pure state case, we have $|\mathbf{b}_\alpha|=\sqrt{2(N-1)/N}$. Considering the vector space $\mathbb{R}^{N^2-1}$, and denoting the $(N^2-1)$-dimensional subspace accessible to the (mixed state) Bloch vector $\mathbf{b}_\alpha$ as \smash{$\Sigma_{\rho_\alpha}^{(N)}$}, one needs to distinguish between two kinds of boundaries, namely its $(N^2-2)$-dimensional {\em topological boundary} \smash{$\partial\Sigma_{\rho_\alpha}^{(N)}$} and its {\em extremal boundary} \smash{$\mathcal{B}_{P_\alpha}^{(N)}$}. The latter is of dimension $2(N-1)$, and it comprises the vectors of maximal $|\mathbf{b}_\alpha|$, \ie it is the space of pure state Bloch vectors, or in other words the \textit{generalized Bloch sphere}, sketched in Fig. \ref{fig:blochsphere}.
\begin{figure*}
	\centering
	\includegraphics[width=\textwidth]{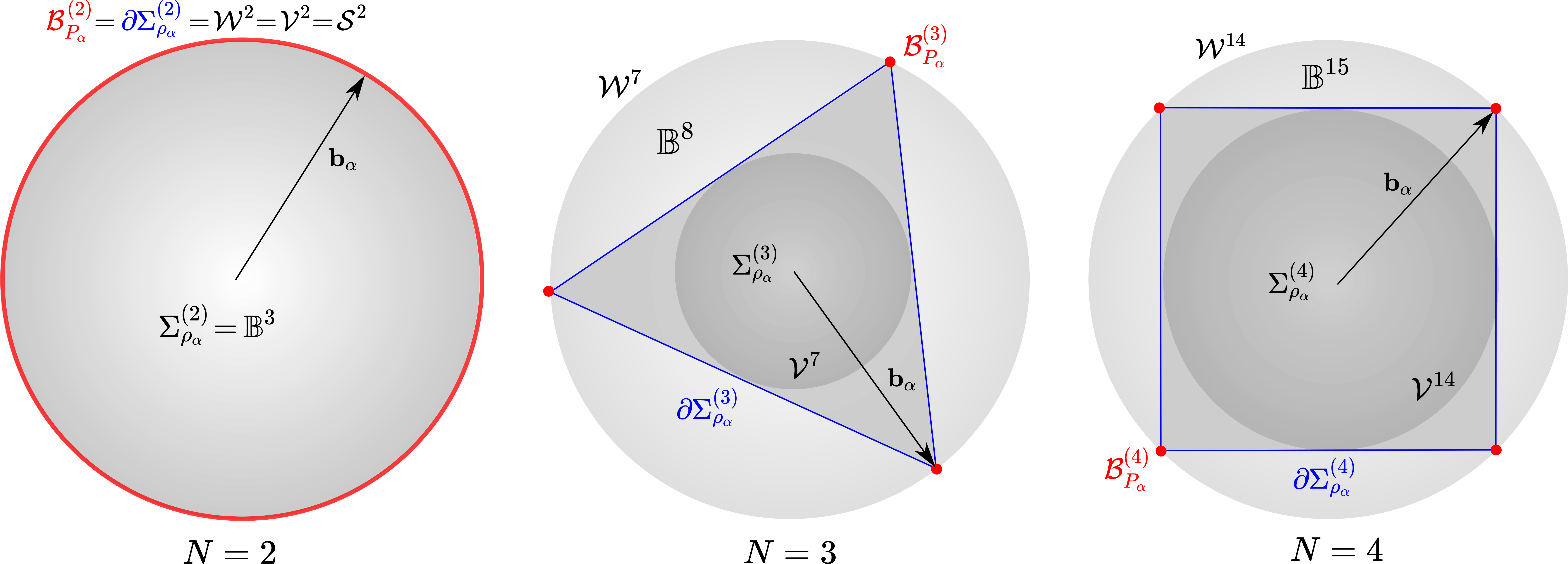}
	\caption{Schematic illustration of the generalized Bloch sphere and other relevant sets, for $N=2,3,4$. The large $(N^2-2)$-sphere \smash{$\mathcal{W}^{N^2-2}$}, of radius $\sqrt{2(N-1)/N}$, corresponds to the highest permissible length of a Bloch vector. In its interior, the Ball \smash{$\mathbb{B}^{N^2-1}$}, there is the space \smash{$\Sigma^{(N)}_{\rho_\alpha}$} accessible to a mixed state of type (\ref{rhoeq}), with topological boundary \smash{$\partial\Sigma^{(N)}_{\rho_\alpha}$}, and in the interior of \smash{$\Sigma^{(N)}_{\rho_\alpha}$} there is the small $(N^2-2)$-sphere \smash{$\mathcal{V}^{N^2-2}$}, of radius $\sqrt{2/[N(N-1)]}$. The extremal boundary \smash{$\mathcal{B}^{(N)}_{P_\alpha}$} of \smash{$\Sigma^{(N)}_{\rho_\alpha}$}, which is the intersection of \smash{$\Sigma^{(N)}_{\rho_\alpha}$} with the large sphere, is the generalized (eigenprojector's) Bloch sphere. For the familiar $N=2$ case, these complications are hidden, since all of the relevant sets coincide.}
	\label{fig:blochsphere}
\end{figure*}

Another set of immediate interest in this discussion is the sphere
\begin{equation}
\mathcal{W}^{N^2-2}\equiv\left\{\mathbf{r}\in\left.\mathbb{R}^{N^2-1}\,\right\vert\,|\mathbf{r}|=\sqrt{\frac{2(N-1)}{N}}\right\}
\end{equation}
that contains all pure state Bloch vectors; in other words, it is the surface of the ball $\mathbb{B}^{N^2-1}$, the smallest ball that contains \smash{$\Sigma^{(N)}_{\rho_\alpha}$}. For $N=2$, one trivially has \smash{$\partial\Sigma^{(2)}_{\rho_\alpha}=\mathcal{B}_{P_\alpha}^{(2)}=\mathcal{W}^2$} (and $\mathcal{W}^2=\mathcal{S}^2$, with the unit two-sphere $\mathcal{S}^2$), as illustrated in Fig. \ref{fig:blochsphere}; but for higher $N$ the three spaces are different. One has \smash{$\mathcal{B}_{P_\alpha}^{(N)}\subset\partial\Sigma^{(N)}_{\rho_\alpha}$} and  \smash{$\mathcal{B}_{P_\alpha}^{(N)}\subset\mathcal{W}^{N^2-2}$}, \ie the generalized Bloch sphere is a proper subset of (i) the boundary of the space of mixed states and (ii) the sphere $\mathcal{W}^{N^2-2}$; in fact, it is their intersection: \smash{$\partial\Sigma^{(N)}_{\rho_\alpha}\cap\mathcal{W}^{N^2-2}=\mathcal{B}_{P_\alpha}^{(N)}$}.
In other words, \smash{if $\mathbf{b}_\alpha$} corresponds to a pure state, it lies on $\mathcal{W}^{N^2-2}$, but not necessarily the other way around, \ie not all points \smash{on $\mathcal{W}^{N^2-2}$} represent physically valid pure states. 

Analogously, the interior of \smash{$\mathcal{W}^{N^2-2}$}, \ie the ball \smash{$\mathbb{B}^{N^2-1}$}, is not composed of only physically valid mixed states. In fact, calculations \cite{Zyczkowski_2003} of the volume of \smash{$\Sigma^{(N)}_{\rho_\alpha}$} show that \smash{$\text{vol}(\Sigma^{(3)}_{\rho_\alpha})/\text{vol}(\mathbb{B}^{8})\approx0.26$} and \smash{$\text{vol}(\Sigma^{(4)}_{\rho_\alpha})/\text{vol}(\mathbb{B}^{15})\approx0.12$}, meaning that most points in \smash{$\mathbb{B}^{N^2-1}$} do actually not represent physically valid states as soon as $N>2$. 

At the origin of all of these complications  is the constraint of positive definiteness, $\rho_\alpha\geq0$: for $N=1$, this condition is trivially fulfilled. For $N=2$, it is equivalent to $\Tr(\rho_\alpha^2)\leq1$, so for a given density matrix there is \emph{one additional constraint} as compared to $N=1$. In the same way, whenever $N$ increases by one, one additional constraint involving traces $D_{\alpha,n}\equiv\Tr(\rho_\alpha^n)$ needs to be fulfilled in order to have positive definiteness.
For example, for $N=3$, there are the three constraints \cite{Kimura_2003} $D_{\alpha,1}=1$, $D_{\alpha,2}\leq1$, and $-2D_{\alpha,3}+3D_{\alpha,2}\leq1$, which can be translated to constraints on the Bloch vectors via Eq. (\ref{rhoeq}). This restrict the elements of $\mathbb{B}^8$ that represent physically valid states.
As a consequence, the relevant sets (including the generalized Bloch sphere) acquire a very nontrivial shape already for $N=3$ \cite{Goyal_2016}:
\begin{equation}
\begin{aligned}
\mathcal{B}^ {(3)}_{P_\alpha}&=\left\{\mathbf{b}_\alpha\in\mathbb{R}^8\left\vert\,|\mathbf{b}_\alpha|^2=\frac{4}{3},\,|\mathbf{b}_\alpha|^2-\mathbf{b}_\alpha\cdot(\mathbf{b}_\alpha\star\mathbf{b}_\alpha)=\frac{4}{9}\right\}\right.,\\
\partial\Sigma^{(3)}_{\rho_\alpha}&=\left\{\mathbf{b}_\alpha\in\mathbb{R}^8\left\vert\,|\mathbf{b}_\alpha|^2\leq\frac{4}{3},\,|\mathbf{b}_\alpha|^2-\mathbf{b}_\alpha\cdot(\mathbf{b}_\alpha\star\mathbf{b}_\alpha)=\frac{4}{9}\right\}\right.,\\
\Sigma^{(3)}_{\rho_\alpha}&=\left\{\mathbf{b}_\alpha\in\mathbb{R}^8\left\vert\,|\mathbf{b}_\alpha|^2\leq\frac{4}{3},\,|\mathbf{b}_\alpha|^2-\mathbf{b}_\alpha\cdot(\mathbf{b}_\alpha\star\mathbf{b}_\alpha)\leq\frac{4}{9}\right\}\right.,
\end{aligned}
\label{B3}
\end{equation}
where the star product is defined in Eq. (\ref{prods}), and the agreement of the first line with Eq. (\ref{unitbloch}) is to be noted.
Similarly, more complicated constraints can be obtained for higher $N$, see Ref. \cite{Kimura_2003}.


Finally, note that there exists another sphere 
\begin{equation}
\mathcal{V}^{N^2-2}\equiv\left\{\mathbf{r}\in\left.\mathbb{R}^{N^2-1}\,\right\vert\,|\mathbf{r}|=\sqrt{\frac{2}{N(N-1)}}\right\}
\end{equation}
inscribed inside \smash{$\Sigma^{(N)}_{\rho_\alpha}$}, \ie \smash{$\partial\Sigma^{(N)}_{\rho_\alpha}$} lies between \smash{$\mathcal{V}^{N^2-2}$ and $\mathcal{W}^{N^2-2}$}. 

The essence of all of these established facts is visualized schematically in Fig. \ref{fig:blochsphere}.
Note that the figure is only a rough sketch supposed to highlight the different sets involved. A visual insight into the true (very complicated) shape of those sets can be gained by considering two- or three-sections, see for instance Refs. \cite{Goyal_2016,Jakobczyk_2001,Kimura_2005,Mendas_2006}.

\section{From eigenstate to eigenprojector picture of the QGT}
\label{AppQGT}

The QGT is a tensor quantifying the overlap of two states that are "infinitesimally close" in parameter space \cite{Provost_1980}: 
$\bra{\psi_\alpha(\mathbf{x})}\ket{\psi_\alpha(\mathbf{x}+d\mathbf{x})}=\mathcal{F}_\alpha(\mathbf{x}) e^{i\phi_\alpha(\mathbf{x})}$, 
where the modulus $\mathcal{F}_\alpha(\mathbf{x})$ has come to be known as \textit{fidelity} \cite{Gu_2010}.
The fidelity is related to the quantum metric, while the phase $\phi_\alpha(\mathbf{x})$ gives rise to the Berry curvature. This can be seen by expanding
\begin{equation}
		\bra{\psi_\alpha(\mathbf{x})}\ket{\psi_\alpha(\mathbf{x}+d\mathbf{x})}=1+\sum_i\bra{\psi_\alpha(\mathbf{x})}\ket{\partial_i\psi_\alpha(\mathbf{x})}dx_i+\frac{1}{2}\sum_{ij}\bra{\psi_\alpha(\mathbf{x})}\ket{\partial_{ij}\psi_\alpha(\mathbf{x})}dx_idx_j+\mathcal{O}(|d\mathbf{x}|^3).
\end{equation}
 Noting that $\Re\bra{\psi_\alpha(\mathbf{x})}\ket{\partial_{ij}\psi_\alpha(\mathbf{x})}=-\Re\bra{\partial_i\psi_\alpha(\mathbf{x})}\ket{\partial_j\psi_\alpha(\mathbf{x})}$ and introducing the Berry connection $\mathcal{A}_{\alpha,i}(\mathbf{x})\equiv i\bra{\psi_\alpha(\mathbf{x})}\ket{\partial_i\psi_\alpha(\mathbf{x})}=-\Im\bra{\psi_\alpha(\mathbf{x})}\ket{\partial_i\psi_\alpha(\mathbf{x})}$, one immediately finds
	\begin{equation}
		\begin{aligned}
			\mathcal{F}_\alpha^2(\mathbf{x})&=[\Re\bra{\psi_\alpha(\mathbf{x})}\ket{\psi_\alpha(\mathbf{x}+d\mathbf{x})}]^2+[\Im\bra{\psi_\alpha(\mathbf{x})}\ket{\psi_\alpha(\mathbf{x}+d\mathbf{x})}]^2\\
			&=1-\sum_{ij}[\Re\bra{\partial_i\psi_\alpha(\mathbf{x})}\ket{\partial_j\psi_\alpha(\mathbf{x})}-\mathcal{A}_{\alpha,i}(\mathbf{x})\mathcal{A}_{\alpha,j}(\mathbf{x})]dx_idx_j+\mathcal{O}(|d\mathbf{x}|^3).
		\end{aligned}
	\end{equation}
Now, the quantum metric tensor $g_{\alpha,ij}(\mathbf{x})$ is defined \cite{Provost_1980, Kolodrubetz_2013} 
by $\sum_{ij}g_{\alpha,ij}dx_idx_j\equiv1-\mathcal{F}_\alpha^2$, such that
\begin{equation}
	g_{\alpha, ij}(\mathbf{x})=\Re\bra{\partial_i\psi_\alpha}\ket{\partial_j\psi_\alpha}+\bra{\psi_\alpha}\ket{\partial_i\psi_\alpha}\bra{\psi_\alpha}\ket{\partial_j\psi_\alpha}.
\end{equation}
Upon calculating the Berry curvature $\Omega_{\alpha,ij}(\mathbf{x})\equiv\partial_i\mathcal{A}_{\alpha,j}-\partial_j\mathcal{A}_{\alpha,i}$, 
one can observe that $g_{\alpha,ij}$ and $\Omega_{\alpha,ij}$ are parts of the single complex tensor (\ref{QG0}).

An eigenstate-based version of the QGT that is used more frequently than Eq. (\ref{QG0}) is given by Eq. (\ref{nextQGT}). From the latter, Eq. (\ref{QG0}) is easily recovered upon using the identity \smash{$\bra{\psi_\alpha}\partial_iH\ket{\psi_\beta}=\bra{\partial_i\psi_\alpha}\ket{\psi_\beta}(E_\alpha-E_\beta)$}, valid for $\alpha\neq\beta$.
To obtain the projector-based formula (\ref{QGT23}), Eq. (\ref{nextQGT}) may be rewritten in an explicitly gauge-invariant form:
\begin{equation}
	T_{\alpha,ij}=\sum_{\beta\neq\alpha}\frac{\Tr\left\{P_\alpha (\partial_iH) P_\beta  (\partial_jH) \right\}}{(E_\alpha-E_\beta)^2},
	\label{nextQGTb}
\end{equation}
where we used the identity $\bra{\psi_\alpha} O \ket{\psi_\alpha}=\Tr\left\{P_\alpha  O P_\alpha \right\}=\Tr\left\{P_\alpha  O \right\}$.
From there, by inserting $H=\sum_\gamma E_\gamma P_\gamma$, it is straightforward to arrive at Eq. (\ref{QGT23}). 

\section{Alternative QGT formula in terms of Bloch vectors}
\label{Appblochalt}

A formula equivalent to Eq. (\ref{geom}) but without derivatives acting on Bloch vectors can be obtained by inserting Eq. (\ref{defproj}) into Eq. (\ref{nextQGTb}) and exploiting the Jacobi identity (\ref{jacobi3}):
\begin{equation}
	\begin{aligned}
		g_{\alpha,ij}&=\sum_{\beta \ne \alpha} \frac{S_{\alpha \beta}^{ij}}{(E_\alpha-E_\beta)^2},
		& \Omega_{\alpha,ij}&=-\sum_{\beta \ne \alpha} \frac{ A_{\alpha \beta}^{ij}}{(E_\alpha-E_\beta)^2}.
		\label{geomb}
	\end{aligned}
\end{equation}
Here, we have defined
	\begin{equation}
		\begin{aligned}
			S_{\alpha \beta}^{ij}&=\frac{4}{N^2}\mathbf{h}^i \cdot \mathbf{h}^j
			+\frac{1}{N} \left[ 
			(\mathbf{b}_\alpha \cdot \mathbf{h}^i)(\mathbf{b}_\beta \cdot \mathbf{h}^j)+(\mathbf{b}_\alpha \cdot \mathbf{h}^j)(\mathbf{b}_\beta \cdot \mathbf{h}^i)\right]\\
			&+\frac{2}{N}(\mathbf{b}_\alpha+ \mathbf{b}_\beta)\cdot(\mathbf{h}^i \star \mathbf{h}^j)+\frac{1}{2}\left[  (\mathbf{b}_\alpha \star \mathbf{h}^i)\cdot(\mathbf{b}_\beta \star \mathbf{h}^j)+(\mathbf{b}_\alpha \star \mathbf{h}^j)\cdot(\mathbf{b}_\beta \star \mathbf{h}^i)\right],\\
			A_{\alpha \beta}^{ij}&=
			\frac{2}{N} (\mathbf{b}_\alpha -\mathbf{b}_\beta)\cdot(\mathbf{h}^i \times \mathbf{h}^j)
			+(\mathbf{b}_\alpha \star \mathbf{h}^i)\cdot(\mathbf{b}_\beta \times \mathbf{h}^j)
			+(\mathbf{b}_\alpha \times \mathbf{h}^i)\cdot(\mathbf{b}_\beta \star \mathbf{h}^j).
			\label{geomc}
		\end{aligned}
	\end{equation}
It is straigthforward to check that \smash{$S_{\alpha \beta}^{ij}$} is symmetric under the exchange of indices $i,j$ (or of $\alpha, \beta$), whereas
\smash{$A_{\alpha \beta}^{ij}$} is antisymmetric under such exchange, and as a consequence $\sum_\alpha \Omega_{\alpha,ij}=0$. 
While apparently much less compact than Eq. (\ref{geom}), the expressions (\ref{geomb}) appear more convenient for numerical implementation.
Conceptually, they mainly distinguish themselves from Eq. (\ref{geom}) in that, on the one hand, they involve 
solely the parametric derivatives of the Hamiltonian vector, while on the other hand they also illustrate the interband nature of the two geometric tensors,
and lastly they also show explicitly the importance of the star product for both the $N$-band quantum metric and Berry curvature.

\section{Berry curvature for $N$-band systems}
\label{AppBerry}

The Berry curvature formula for arbitrary SU($N$) systems follows from combining Eqs. (\ref{Projectormain2}), (\ref{etan}) \& (\ref{geom}), together with the total antisymmetry of the triple product $\mathbf{m}\cdot(\mathbf{n}\times\mathbf{o})=f_{abc}n_ao_bm_c$ and the orthogonality relations (\ref{ortho}) or (\ref{orth}):
\begin{equation}
\Omega_{\alpha,ij}=-\frac{2\,\mathbf{b}_\alpha}{[q_N'(E_\alpha)]^2}\cdot\left(\sum_{n,m=1}^{N-1}q_{N-1-n}(E_\alpha)q_{N-1-m}(E_\alpha)\,\boldsymbol{\tau}_n^{(i)}\times\boldsymbol{\tau}_m^{(j)}\right).
\label{be}
\end{equation}
Here we use the polynomials $q_n(z)$ and have defined vectors
\begin{equation}
\boldsymbol{\tau}_n^{(i)}\equiv\frac{1}{N}\sum_{p=0} ^{n-1} C_p\partial_i \mathbf{h}_{\star} ^{(n-1-p)}.
\label{ci}
\end{equation}
With the notation $\mathbf{m}^i\equiv\partial_i\mathbf{m}$, the first few \smash{$\boldsymbol{\tau}_n^{(i)}$} read
\begin{equation}
\begin{aligned}
\boldsymbol{\tau}_1^{(i)}&=\mathbf{h}^i,&&& \boldsymbol{\tau}_2^{(i)}&=\mathbf{h}_\star^i,&&&
\boldsymbol{\tau}_3^{(i)}&=\frac{C_2}{N}\mathbf{h}^i+\mathbf{h}_{\star\star}^i.
\end{aligned}
\end{equation}
From Eq. (\ref{be}), one can obtain the Berry curvature for any given $N$. 

For $N=2$, Eq. (\ref{Q2}) is immediately obtained.
For $N=3$, analogously, we have
\begin{equation}
\begin{aligned}
\Omega_{\alpha,ij}&=-\frac{4(E_\alpha\mathbf{h}+\mathbf{h}_\star)}{(3E_\alpha^2-\frac{C_2}{2})^3}\cdot\left[E_\alpha^2\,\mathbf{h}^i\times\mathbf{h}^j+E_\alpha\left(\mathbf{h}^i\times\mathbf{h}_\star^j+\mathbf{h}_\star^i\times\mathbf{h}^j\right)+\,\mathbf{h}_\star^i\times\mathbf{h}_\star^j\right],
\end{aligned}
\label{om32}
\end{equation}
or equivalently Eq. (\ref{enbar}) in the main text. This result can be simplified due to the following identities:
\begin{equation}
\begin{aligned}
\mathbf{h}\cdot(\mathbf{h}^i\times\mathbf{h}_\star^j)&=\mathbf{h}\cdot(\mathbf{h}_\star^i\times\mathbf{h}^j)=\mathbf{h}_\star\cdot(\mathbf{h}^i\times\mathbf{h}^j),\\
\mathbf{h}_\star\cdot(\mathbf{h}^i\times\mathbf{h}_\star^j)&=\mathbf{h}_\star\cdot(\mathbf{h}_\star^i\times\mathbf{h}^j)=\mathbf{h}\cdot(\mathbf{h}_\star^i\times\mathbf{h}_\star^j),\\
\mathbf{h}_\star\cdot(\mathbf{h}_\star^i\times\mathbf{h}_\star^j)&=\left[-\frac{2}{3}(\mathbf{h}\cdot\mathbf{h}_\star)\mathbf{h}+|\mathbf{h}|^2\mathbf{h}_\star\right]\cdot(\mathbf{h}^i\times\mathbf{h}^j).
\end{aligned}
\label{firstids}
\end{equation}
The first two lines are valid for general $N$ and can be proved using the second Jacobi identity (\ref{jacobi2}). The proof of the third line requires the Jacobi indentity as well, but additionally the SU($3$)-specific identities (\ref{starrelsb}).
Inserting all of these identities into Eq. (\ref{om32}), and exploiting the characteristic equation $E_\alpha^3=\frac{C_2}{2}E_\alpha+\frac{C_3}{3}$, one obtains the generic SU($3$) Berry curvature formula
\begin{equation}
	\begin{aligned}
		\Omega_{\alpha,ij}=\frac{-4}{\left(3E_\alpha^2-|\mathbf{h}|^2\right)^3}&\left\{E_\alpha\left[|\mathbf{h}|^2\mathbf{h}\cdot\left(\mathbf{h}^i\times\mathbf{h}^j\right)
		+3\,\mathbf{h}\cdot\left(\mathbf{h}_\star^i\times\mathbf{h}_\star^j\right)\right]+\left(3E_\alpha^2+|\mathbf{h}|^2\right)\mathbf{h}_\star\cdot\left(\mathbf{h}^i\times\mathbf{h}^j\right)\right\}.
		\label{om3}
	\end{aligned}
\end{equation}
The advantage of this formula, as compared to Eq. (\ref{enbar}), consists in the fact that it contains only three terms, and that the Berry curvature sum rule $\sum_\alpha\Omega_{\alpha,ij}=0$ is more evident, since $\sum_\alpha E_\alpha(3E_\alpha^2-|\mathbf{h}|^2)^{-3}=0$ and
$\sum_\alpha (3E_\alpha^2+|\mathbf{h}|^2)(3E_\alpha^2-|\mathbf{h}|^2)^{-3}=0$.

In principle, one can continue in this way to obtain the Berry curvature for any $N$. As a shortcut, it is however useful to realize that, when the Bloch vectors (\ref{balpha34}) are substituted into Eq. (\ref{geom}), the derivatives effectively do not act on the prefactors but only on the vectors $\mathbf{h}$, $\mathbf{h}_\star$, \etc~ This is due to the orthogonality relations (\ref{ortho}) \& (\ref{orth}). For example, for the $N=2$ case, it suffices to replace $\mathbf{b}_\alpha^i\rightarrow\frac{1}{E_\alpha}\mathbf{h}^i$ in Eq. (\ref{geom}), which allows to directly read off the Berry curvature: $\Omega_{\alpha,ij}=-\frac{1}{2}\frac{\mathbf{h}}{E_\alpha}\cdot(\frac{\mathbf{h}^i}{E_\alpha}\times\frac{\mathbf{h}^j}{E_\alpha})$. Similarly, for $N=3$, it suffices to replace $\mathbf{b}_\alpha^i\rightarrow\frac{2}{3E_\alpha^2-C_2/2}(E_\alpha\mathbf{h}^i+\mathbf{h}_\star^i)$, and so on for higher $N$. In this way, one directly obtains Eqs. (\ref{enbar})--(\ref{om5}) in the main text.

\end{widetext}

\bibliography{ref}

\end{document}